# A Combined Computational and Experimental Investigation on Evaporation of a Sessile Water Droplet on a Heated Hydrophilic Substrate

Manish Kumar, Rajneesh Bhardwaj[*]

Department of Mechanical Engineering,

Indian Institute of Technology Bombay, Mumbai 400076, India

[*]Corresponding author (Email: rajneesh.bhardwaj@iitb.ac.in)

Phone: +91 22 2576 7534

*Abstract*

We numerically and experimentally investigate evaporation of a sessile droplet on a heated substrate. We develop a finite element (FE) model in two-dimensional axisymmetric coordinates to solve coupled transport of heat in the droplet and substrate, and of the mass of liquid vapor in surrounding ambient while assuming diffusion-limited, quasi-steady evaporation of the droplet. The two-way coupling is implemented using an iterative scheme and under-relaxation is used to ensure numerical stability. The FE model is validated against the published spatial profile of the evaporation mass flux and temperature of the liquid-gas interface. We discuss cases in which the two-way coupling is significantly accurate than the one-way coupling. In experiments, we visualized side view of an evaporating microliter water droplet using a high-speed camera at different substrate temperatures and recorded temperature of the liquid-gas interface from the top using an infrared camera. We examine the dependency of inversion of the temperature profile across the liquid-gas interface on the ratio of the substrate thickness to the wetted radius, the ratio of the thermal conductivity of the substrate to that of the droplet and contact angle. A regime map is plotted to demarcate the inversion of the temperature profile for a wide range of these variables. A comparison of measured evaporation mass rate with the computed values at different substrate temperature show that the evaporation mass rate increases non-linearly with respect to the substrate temperature, and FE model predicts these values close to the experimental data. Comparisons of time-averaged evaporation mass rate obtained by the previous and present models against the measurements suggest that the evaporative cooling at the interface and variation of diffusion coefficient with temperature should be taken into account in the model in order to accurately capture the measurements. We compare the measurements of time-varying droplet dimensions and of temperature profile across the liquid-gas interface with the numerical results and found good agreements. We quantify increase in the evaporation mass flux and evaporation mas rate by the substrate heating and present the combined effect of substrate heating, the ratio of the substrate thickness to the wetted radius, substrate-droplet thermal conductivity ratio and the contact angle on the evaporation mass rate.

# 1 Introduction

In the last decade, evaporation of a sessile, pure liquid droplet on a solid surface is a much-studied problem, owing to several technical applications such as evaporative spray cooling and inkjet printing, etc. The physics involved during the evaporation is briefly described as follows. In the absence of any external convection, the evaporation occurs by diffusion of liquid vapor in surrounding gas. The evaporation mass flux ($j$) [kg/m$^2$-s] on the liquid-gas interface is non-uniform, and the largest evaporation near the contact line generates evaporative-driven radially outward flow inside the droplet [1]. Heat transfer occurs mostly by conduction in the droplet and substrate and the non-uniform evaporative flux also results in non-uniform cooling at the liquid-gas interface by latent heat of evaporation. Depending upon the roughness of the substrate, the contact line may remain pinned or may recede at a constant contact angle during the evaporation.

Several previous theoretical and numerical studies investigate the evaporation of a sessile droplet on a *non-heated* substrate. Deegan [1] and Hu and Larson [2] reported simplified expressions of *j* valid for contact angles 0 to 90º for quasi-steady-state evaporation of a spherical cap drop with a pinned contact line on a substrate kept at ambient temperature. In a follow-up study, Hu and Larson [3] derived an analytical expression of velocity field inside an evaporating sessile droplet using lubrication theory. Popov [4] analytically solved the vapor concentration field using toroidal coordinates and gave expressions of evaporation mass flux, evaporation mass rate and evaporation time, valid for any arbitrary contact angle. The sign of temperature gradient along the liquid-gas interface determines the direction of thermocapillary (or Marangoni) flow inside the droplet and is influenced by contact angle [5], the ratio of thermal conductivity of the substrate to that of droplet [6] and the ratio of substrate thickness to wetted radius [7].

In the context of theoretical studies for *heated* substrates, Sobac and Brutin [8] modeled the droplet evaporation by extending the model of Hu and Larson and by considering the temperature of the liquid-gas interface equal to substrate temperature, thereby ignoring the heat transfer in droplet and substrate. In this study [8], comparisons of the model predictions with measurements showed that the heat transfer should be considered in the model larger substrate heating, in order to accurately capture the measurements. Zhang et al. [9] solved energy equation in the droplet and substrate in axisymmetric coordinates assuming quasi-steady evaporation. They used spatial profile of evaporation mass flux described by Hu and Larson [2] in their model and showed that the spatial profile of temperature at the liquid-gas interface depends on the ratio of thermal conductivity of the substrate to that of the droplet. Simulations by Barmi and Meinhart [10] quantified the internal convection against Marangoni number and they found that the Marangoni convection becomes negligible as the droplet volume decreases during the evaporation. Maatar et al. [11] numerically investigated the evaporation of water volatile liquid droplets considering transient effects while modeling the energy equation in the droplet and the substrate. They showed that the transient effects are important to consider in the model for a thicker substrate with lower thermal diffusivity. Xu and Ma [12] proposed a "combined field approach" to couple the energy equation and Laplace equation of diffusion of vapor concentration. By assuming a linear variation of saturated concentration with temperature, they proposed a unified way to solve the two governing equations. This method does not need iterations between the two governing equations. In a follow-up paper, Wang. et al. [13] extended the model of Xu and Ma [12] to investigate the combined effect of evaporative cooling, and thickness and thermal conductivity of the substrate. They showed that for larger evaporative cooling, the influence of substrate is significant. Liu et al. [14] numerically showed that the transient effects during the evaporation of volatile droplets are important and assumption of quasi-steady evaporation is not valid in such cases. Very recently, Bouchenna et al. [15] proposed a model to study the flow inside the evaporating water droplet on a heated substrate and showed the existence of multicellular flow pattern at smaller contact angles and larger substrate heating.

In the context of recent experimental studies, David et al. [16] experimentally investigated the effect of substrate thermal conductivity and concluded that it influences the evaporation mass rate. In particular, significant evaporating cooling can occur by an insulating substrate. Bhardwaj et al. [17] recorded impact and evaporation of an isopropanol droplet using high-speed visualization. They used a laser-based thermos-reflectance method to measure liquid-solid interface temperature and showed that the temperature increases exponentially during the impact and it undergoes a slight linear decrease during the evaporation. Ghasemi and Ward [18] experimentally showed that the thermocapillary convection is the dominant mode of heat transfer near the contact line, however, heat conduction dominates at the apex of the droplet. Sobac and Brutin [8] experimentally investigated the thermal effects of the substrate by recording an evaporating water droplet on hydrophilic and hydrophobic engineered aluminum substrates. Lopes et al. [19] investigated the effect of thermal properties of the substrate on evaporation time of a sessile droplet and found that evaporation accelerates on a substrate with larger thermal conductivity. Very recently, Bazargan and Stoeber [20] experimentally investigated the effect of substrate conductivity on evaporation of water droplets of 100-500 μm diameter. The comparison of these measurements with a one-dimensional heat transfer model showed the existence of a critical radius of sessile droplet below which substrate cooling effects the total evaporation time.

Several recent studies have reported the measurement of liquid-gas interface temperature using infrared thermography. Brutin et al. [21] visualized thermal-convective instabilities during evaporation of droplets of volatile liquids on a heated surface using infrared visualization. Fabien et al. [22] recorded temperature of the liquid-gas interface of an evaporating water droplet on heated substrates using infrared thermography and plotted the temporal evolution of the temperature in different cases of substrate temperatures. Very recently, Fukatani et al. [23] recorded hydrothermal waves in an evaporating ethanol droplet using infrared thermography and showed that these waves can be influenced by relative humidity.

Most of the previous models of evaporating sessile droplet on non-heated or heated substrates were based on the following assumptions: heat transfer is only in the axial direction [7], liquid-gas interface temperature is equal to substrate temperature [8] and saturated liquid-vapor concentration varies linearly with temperature [12, 13]. The expressions of evaporation mass rate and spatial variation of evaporative flux reported in previous studies [2, 4] are valid for a substrate at ambient temperature. In addition, the combined effect of parameters, namely, substrate heating, substrate thickness, contact angle, thermal properties of droplet and substrate, substrate heating have not been reported before. To this end, we present a combined numerical and experimental study with the following objectives. First, we develop and validate a model for droplet evaporation on a heated substrate which solves coupled energy and mass transport equation in order to account for heat transfer in droplet as well substrate. Second, using the model, we investigate the effect of geometry and thermophysical properties of droplet and substrate, and substrate temperature on evaporation characteristics. Third, we perform experiments to measure time-varying droplet shapes using high-speed visualization and temperature of the liquid-gas interface using infrared thermography.

Finally, the measurements are compared with the model predictions in order to investigate the fidelity of the model and understand the coupled physics.

## 2 Computational Model

We extend the models reported in previous studies [2, 8, 12, 13] to account the evaporation of a sessile droplet on a heated hydrophilic substrate. In particular, we develop a two-way coupling of the energy equation in droplet and substrate and transport of liquid vapor outside the droplet. The definitions of the notations used in the following sections are listed in Table 1.

### 2.1 Governing equations and boundary conditions

We consider diffusion-limited, quasi-steady evaporation of a sessile droplet with a pinned contact line on a heated hydrophilic substrate. The wetted diameter of the droplet is taken lesser than capillary length so that the droplet maintains a spherical cap shape throughout the evaporation. The validity of the quasi-steady evaporation is examined by considering the ratio of heat equilibrium time in a droplet ($t_h$) and its total evaporation time ($t_F$), as discussed by Larson [24] and is given by,

$$\frac{t_h}{t_F} \sim 5 \frac{D}{\alpha_d} \frac{h_d}{R} \frac{c_{sat}}{\rho_d} \tag{1}$$

where $D$, $\alpha_d$, $h_d$, $R$, $c_{sat}$ and $\rho_d$ are diffusion coefficient of liquid vapor in the air, the thermal diffusivity of the droplet, height of the droplet, the wetted radius of the droplet, saturation vapor concentration and density of droplet, respectively. We plot $t_h/t_F$ for several liquids with respect to temperature in Figure 1. The properties of liquids used in this figure are given in supplementary information. The values of $D$, $\alpha_d$, $h_d/R$ and $\rho_d$ to calculate $t_h/t_F$ for water are $2.4 \times 10^{-5}$ m²/s, $1.45 \times 10^{-7}$ m²/s, 0.36 and 997 kg/m³, respectively. A horizontal dashed line representing $t_h/t_F = 0.1$ shows the limit of quasi-steady evaporation. As seen from the plot, water droplets with average temperature ($T_{avg}$) of lower than around 75°C satisfy the assumption of quasi-steady evaporation because of $t_h/t_F < 0.1$ for $T_{avg} < 75$°C. The corresponding value of $T_{avg}$ is significantly lower for alcohols in Figure 1. We consider microliter water droplets in the present study and choose substrate temperature such that the assumption of quasi-steady evaporation is valid.

We neglect thermocapillary or thermal convection inside the droplet and this is justified for low Péclet number (*Pe*). We estimate *Pe* based on the Marangoni flow velocity ($V_{Ma}$) for cases of the microliter water droplets on heated substrate considered in this study. $V_{Ma}$ is calculated using the analytical expression suggested in Refs. [25, 26] and the temperature gradient is obtained from the measurements in section 4.2.2 (discussed later). The estimated range of *Pe* is from 1 to 24 and as suggested by Larson [24] heat convection is significant if *Pe* exceeds around 10. We also note that the Marangoni convection or heat convection dominates heat conduction only near the contact line [6, 18]. Therefore, the assumption of neglecting the convection in the model is valid for the water droplets and intensity of substrate heating considered in this study.

With above assumptions, the energy equation in the droplet (*i* = *d*) and the substrate (*i* = *s*) for the temperature field simplifies to,

$$\nabla^2 T_i = 0 \qquad (2)$$

A perfect thermal contact between the drop and the substrate is assumed. The computational domain and boundary conditions for eq. (2) are shown in Figure 2(a). Neumann boundary condition for temperature is applied to the top horizontal surface of the substrate, right boundary of the substrate and at *r* = 0. Along the bottom boundary of the substrate, a constant temperature boundary condition (*T* = $T_S$) is applied and jump energy boundary condition is applied along the liquid-gas interface of the droplet

$$jL = -k\nabla T \cdot \mathbf{n} \qquad (3)$$

where *j*, *L*, *k*, and **n** are evaporative flux [kg/m$^2$-s] at the liquid-gas interface of the droplet, latent heat of the evaporation [J/kg], the thermal conductivity of the liquid [W/m-K] and unit normal vector, respectively.

In absence of external convection, the diffusion of liquid vapor in surrounding gas concentration (*c*) is governing by the following Laplace equation,

$$\nabla^2 c = 0 \qquad (4)$$

The boundary conditions for eq. (4) are described in Figure 2(a). Neumann boundary condition for *c* is applied for *r* > *R*, *z* = 0 and *r* = 0. The following Dirichlet boundary conditions are applied in far-field *r* = ∞, *z* = ∞ and at the liquid-gas interface respectively, *c* = $Hc_\infty$ and *c* = $c_{LG}$, where $c_{LG}$ is the saturated concentration [kg/m$^3$] of the liquid-vapor near the interface and $c_\infty$ is the saturated concentration of liquid vapor in the far-field (ambient), and *H* is the relative humidity. A domain size independence study shows that it is sufficient to consider the far field ($r = \infty$, $z = \infty$) at $r = 50R$, $z = 50R$ (Figure 2). The evaporative flux $j$ at the liquid-gas interface is expressed as follows [3, 4]:

$$j(r,T) = -D(T)[\nabla c . \mathbf{n}]\big|_{LG} \tag{5}$$

where diffusion coefficient of liquid vapor in gas ($D$) is a function of temperature ($T$),

We consider evaporating water droplet in the air on a heated glass substrate and the thermophysical properties of water and glass used in the model are listed in Table 2. The saturated concentrations at a given temperature for water vapor are obtained using the following third order polynomial [27]:

$$c_{sat} = [9.99 \times 10^{-7} T^3 - 6.94 \times 10^{-5} T^2 + 3.20 \times 10^{-3} T - 2.87 \times 10^{-2}] \tag{6}$$

where $T$ is the temperature in °C. The dependence of diffusion coefficient of water vapor on temperature (°C) is given by [27]:

$$D(T) = 2.5 \times 10^{-4} \exp\left(-\frac{684.15}{T + 273.15}\right) \tag{7}$$

## 2.2 Numerical algorithm and methodology

We employ Galerkin finite-element method to solve coupled eq. (2) and (4) for temperature and vapor concentration field in axisymmetric cylindrical coordinates. Previous studies (for example, Ref. [9]) employed *one-way coupling* in which the vapor concentration field is obtained by solving eq. (4) assuming liquid-gas interface temperature equal to the substrate temperature and evaporation mass flux is calculated using eq. (6). The temperature field is obtained by solving eq. (2) using $j$ thus obtained. In the present work, we implement *two-way coupling* between the eq. (2) and (4) using an iterative approach. We compare the results of one-way and two-way coupling in section 2.5. The model is implemented in MATLAB© and the functions used in the implementation are given in supplementary information. The algorithm of the two-way coupling is described as follows. In a given i[th] iterative step, the following calculations are performed:

1) Solve energy equation (eq. (2)) with jump energy boundary condition (eq. (3)) with $j$ obtained in the i-1 iteration.
2) Solve vapor concentration equation (eq. (4)) with prescribed vapor concentration based on the temperature at the liquid-gas interface obtained in step 1.

3) Calculate evaporation mass flux $j_i$ using eq. (5) based on vapor concentration field obtained in step 2 and diffusion coefficient (eq. (7)) based on the temperature at the liquid-gas interface obtained in step 1.
4) Underrelax $j_i$ obtained in step 3 to obtain $j_{i,\ revised}$.
5) Repeat steps 1 to 4 until the following convergence criterion is met: $L_2$-norm of the temperature field is lesser than $10^{-3}$.

## 2.3 Grid size convergence study

In order to achieve grid-size convergence, we test five grids and number of nodes in the computational domain of droplet-substrate and ambient are listed in Table 3. The following parameters are used for these simulations: the ratio of substrate thickness to the wetted radius, $h_S/R = 10$; contact angle $\theta = 20°$, the ratio of thermal conductivity of the substrate to that of the droplet, $k_S/k_d = 1.58$. We calculate the $L_2$-norm error ($\varepsilon$) for a scalar field $\psi$ (temperature or concentration) with respect to the most refined respective grid and is expressed as follows:

$$\varepsilon = \left[ \frac{1}{N^2} \sum_{i=1}^{N} \left( \psi_i^N - \psi_i^{N_{max}} \right)^2 \right]^{1/2} \tag{8}$$

where $N$ is number of nodes in the grid, $N_{max}$ is number of nodes in the most refined grid (Grid 5 in Table 3). Note that $\psi_i^{N_{max}}$ is calculated by the linear interpolation on the most refined grid at the same location of $\psi_i^N$. Figure 3 shows the $L_2$-norm error for temperature ($T$) and vapor concentration ($c$) field against the different grid tested on a log-log plot and a line with slope = 2 is also plotted for assessing the order of accuracy of the solver. We conclude that our solutions are second-order accurate and grid independent. The results of grid 4 and grid 5 are very close (Figure 3) and we employ grid 5 for the simulations presented in the paper.

## 2.4 Code validations

In order to validate the present model, we compare computed the spatial variation of evaporation mass flux and temperature at the liquid-gas interface with previously reported results. First, we compare the variation of non-dimensional temperature at the liquid-gas interface ( $\hat{T}_{LG} = (T - T_{ref})/(T_s - T_{ref})$) with non-dimensional radial distance, $\hat{r} = r/R$ with published results.

^ denotes the non-dimensional value and $T_s$ is substrate temperature. $T_{ref}$ is reference temperature and is taken as 0 °C. Figure 4(a) shows the comparison of $\hat{T}_{LG}$ simulated in present work and corresponding numerical data of Zhang et al. [9] for different contact angles ($\theta$). The substrate-droplet thermal conductivity ratio ($k_S/k_d$) and substrate thickness-wetted radius ratio ($h_S/R$) are 10 and 0.3, respectively. The temperature of the substrate ($T_S$) is 22°C. The agreement with published results is good, with a maximum error of 1.3%. In Figure 4(b), the comparisons are presented for the low values of $k_S/k_d$ and $h_S/R$. These values are 1.58 and 0.15, respectively. The comparisons are in good agreement with a maximum error of around 4%. In this case, the temperature of the top of the droplet is larger than that near the contact line because the cooling due to the latent heat of evaporation dominates over thermal energy available from the substrate.

Third, we compare the variation of computed non-dimensional evaporation mass flux ($\hat{j}$) with non-dimensional radial distance, $\hat{r} = r/R$, with the corresponding variation of analytical expression reported by Hu and Larson [2]. We use the same expression of $\hat{j}$ as given by Wang et al. [13] and is defined as follows,

$$\hat{j} = \frac{jR}{D_\infty c_\infty (1-H)} \qquad (9)$$

where $D_\infty$ and $c_\infty$ are diffusion coefficient and saturation vapor concentration at ambient temperature, respectively. $H$ is relative humidity of the ambient. The expression of evaporative flux on the liquid-gas interface ($j$) is expressed as a function of non-dimensional radial distance is given by [2]:

$$j(\hat{r}) = \frac{Dc_{sat}(1-H)}{R}(0.27\theta^2 + 1.30)(0.6381 - 0.2239(\theta - \pi/4)^2)(1-\hat{r}^2)^{-\lambda(\theta)} \qquad (10)$$

where $\theta$ is wetting angle and $\lambda(\theta) = 0.5 - \theta/\pi$. Figure 5(a) shows the comparison of the profile of $\hat{j}$ simulated for different cases of contact angle. The substrate-droplet thermal conductivity ratio ($k_S/k_d$) and substrate thickness-wetted radius ratio ($h_S/R$) are 10 and 0.3, respectively in these simulations. The results of present work show good agreement with the model of Hu and Larson with a maximum error of 2.6%. Finally, Figure 5(b) shows the comparison of $\hat{j}$ with simulations of Wang et al. [13] for substrates of different thermal conductivity. The present results are in good agreement with a maximum error of 5.7%.

## 2.5 Comparison between two-way and one-way coupling

In order to examine the fidelity of the two coupling schemes, we perform simulations using these schemes for the following parameters, $\theta = [20°, 90°]$ at $T_S = [25°C, 90°C]$. This comparison takes the case, in which, water droplet is resting on a thin substrate ($h_S/R = 0.3$) with substrate-droplet conductivity ratio ($k_S/k_d$) of 10 and evaporating at ambient conditions, i.e., ambient temperature 25°C and humidity 40%.

Figure 6(a) shows the comparison of $\hat{j}$ computed by the one-way and two-way coupling at $T_S = 25°C$. At smaller contact angle, $\theta = 20°$, $\hat{j}$ predicted by one-way coupling model matches closely with two-way coupling with a difference of 4.5% with respect to two-way coupling at $r/R = 0$, however, at higher contact angle, $\theta = 90°$, it does not match and overpredicts it with a difference of 32.5% with respect to two-way coupling at $r/R = 0$. Therefore, at higher contact angle, thermal resistance for heat flow from the substrate to the liquid-gas interface of the droplet is larger, which reduced available thermal energy from the substrate, resulting in significant evaporation cooling. In one-way coupling, this effect is not captured and hence, it overpredicts $\hat{j}$ because the calculation of $\hat{j}$ assumes $T_{LG} = T_S$.

Figure 6(b) shows $T_{LG}$ predicted by the one-way and two-way coupling model at $T_S = 25°C$. Figure 6(b) shows that at lower contact angle, $\theta = 20°$, the temperature drop from the edge ($r/R = 1$) of the droplet to center ($r/R = 0$) of the droplet ($\Delta T = T_{edge} - T_{top}$) is very low (0.2°C), computed by both coupling schemes. However, in case of higher contact angle, $\theta = 90°$, the temperature difference between droplet edge and center is more than 1°C, computed by both coupling schemes. $\Delta T$ in one-way coupling is 26.6% lower as compared to two-way coupling. This is explained by the fact that one-way coupling overpredicts $\hat{j}$ based on $T_S$ (Figure 6(a)) to compute the evaporative cooling.

We note similar characteristics in case of $T_S = 90°C$. We also plot $\hat{j}$ and $T_{LG}$ at $T_S = 90°C$ in Figure 6(c) and (d), respectively. $T_{LG}$ and $\hat{j}$ predicted by one-way coupling model at the $r/R = 0$ for $\theta = 90°$ are 45% lower and 200% larger than predicted by two-way coupling, respectively. Therefore, the two-way coupling is more accurate for larger $\theta$, that corresponds to significant evaporative cooling at liquid-gas interface. Similarly, this coupling scheme is also accurate for a substrate with smaller thermal conductivity and larger thickness because these conditions correspond to large cooling at the interface. Since the substrate heating amplifies the error in the one-way coupling for a given set of parameters, the two-way coupling is more accurate for a heated substrate.

## 2.6 Expression of evaporation mass rate

The time-varying droplet volume is calculated using the same method proposed by Hu and Larson [2] and is described as follows. The moving liquid-gas interface is considered a series of solutions of the coupled system of equations. The evaporated volume in a time-step is calculated by integrating the evaporation mass flux over the liquid-gas interface and multiplying the integral with the time step. Using the remaining volume, a revised droplet shape is defined by assuming it as a spherical cap. These steps are repeated until the end of evaporation and the time step is used as, $0.02 t_F$, where $t_F$ is the total evaporation time. The instantaneous evaporation mass rate is non-dimensionalized as follows,

$$\hat{m} = \frac{1}{RD_\infty c_\infty (1-H)} \int_A j(r) dA \qquad (11)$$

where evaporation mass flux $j(r)$ is integrated over the liquid-gas interface of area $A$. $D_\infty$ and $c_\infty$ are the diffusion coefficient and saturation vapor concentration at ambient temperature, respectively. The time-averaged, non-dimensional evaporation mass rate is defined by Sobac and Brutin [8] and is defined as follows,

$$\hat{m}_{avg} = \frac{1}{t_F} \int_0^{t_F} \hat{m} dt \qquad (12)$$

## 3 Experimental details

A schematic of the experimental setup used in the present study is shown in Figure 7(a). 2.0±0.15 μL water droplets were generated using a micropipette (Prime, Biosystem Diagnostic Inc., India) and were gently deposited on a glass slide (Borosil Inc, India) of dimensions 75 × 25 × 1 mm. The slides were sequentially cleaned with isopropanol and deionized water and were allowed to dry out completely in the ambient air. The slides were heated using a hot plate at different temperatures listed in Table 4. The droplets were deposited on the slide after the desired temperature was attained and remained steady on the surface at least for 4-5 minutes.

As shown in Figure 7(a), the time-varying shapes of evaporating droplet were visualized from the side using a high-speed camera (MotionPro, Y-3 classic) with long distance working objective (Qioptiq Inc.). A white LED lamp was used as a backlight source. The working distance, image resolution, and pixel resolution are 9.5 cm, 600 × 450 and 71 pixels/mm, respectively. The images were recorded at 10 and 100 frames per second for the non-heated and heated substrate, respectively. A higher value of frames per second was used in the latter case due to faster evaporation.

We used a high-speed infrared camera (A6703sc, FLIR Systems Inc.) with 25 mm (f/2.5) IR lens and close focusing extender ring to record transient temperature of the liquid-gas interface. The infrared camera was mounted to capture the top view of the droplet. The working distance, image resolution, and pixel resolution are 10.6 cm, 300 × 256 and 14 pixels/mm, respectively. The frame rate is same as mentioned above for the high-speed camera. The emissivity of water and glass are taken as 0.95. The measured temperature is close to liquid-gas interface temperature because water is opaque to the infrared radiation. The calibration of the infrared camera was reported our previous study [26] and the uncertainty in the temperature response of IR camera is around ±1.0°C.

All the experiments were performed in a controlled environment at 24±2°C and 35±4% relative humidity. Figure 7(b) shows a typical droplet shape obtained using high-speed visualization and different droplet dimensions namely contact angle ($\theta$), droplet height ($h_d$) and wetted diameter ($d$) are shown in this figure. $h_d$ and $d$ are obtained after standard image processing techniques. In order to calculate the volume of the evaporating droplet, we assume it to be a spherical cap. The volume ($V$) and contact angle ($\theta$) are expressed as follows,

$$V = \pi h_d (3R^2 + h_d^2)/6, \ \theta = 2\tan^{-1}(h_d/R) \tag{13}$$

where $R$ is the wetted radius ($R = d/2$). The initial wetted radius and initial equilibrium contact angle of the droplet measured at different temperatures of the heated glass are listed in Table 4. The uncertainty in contact angle and wetted radius measurements are ±2° and ±0.05 mm, respectively.

## 4   Results and Discussions

First, numerical results on a non-heated (or isothermal) substrate are presented (section 4.1). Second, we present the comparisons between numerical and experimental results (section 4.2). Finally, we present the numerical results of evaporation mass flux and evaporation mass rate on a heated substrate (section 4.3).

### 4.1   Inversion of temperature gradient across the liquid-gas interface on an isothermal substrate

It has been well-established that the ratio of thermal conductivity of the substrate to that of the droplet ($k_S/k_d$) controls the temperature gradient across the liquid-gas interface [27]. As shown schematically in Figure 8(a), $\Delta T > 0$ induces a thermal Marangoni flow from the edge towards the center of the droplet. By contrast, $\Delta T < 0$ induces a flow in the opposite direction. At larger $k_S/k_d$, heat supplied near the contact line is larger than the top of the droplet, which results in a positive thermal gradient across the interface ($\Delta T > 0$, as shown in Figure 8(a), left). By contrast, lower $k_S/k_d$ corresponds to $\Delta T < 0$, as shown in Figure 8(a), right. In addition, the sign of $\Delta T$ also depends upon the contact angle ($\theta$) and $\Delta T < 0$ occurs at smaller $\theta$ [5]. Recently, Liu et al. [14] showed that the sign of $\Delta T$ is controlled by varying ratio of substrate thickness to the wetted radius of the droplet ($h_S/R$).

To examine the combined effect of $h_S/R$, $k_S/k_d$ and $\theta$ on the sign of $\Delta T$, we performed several simulations. The following range of the parameters are considered in simulations: $h_S/R$ = [0.1, 0.8], $k_S/k_d$ = [0.37, 4.2] and $\theta$ = [5º, 30º]. The frames of Figure 8(b) show contours of computed $\Delta T$ on $k_S/k_d$ - $\theta$ plane for different values of $h_S/R$. In addition to contours, two curves of $\Delta T = 0$ are plotted in each frame of Figure 8(b). A thick black curve and a broken curve are obtained from present two-dimensional (2D) simulation and one-dimensional (1D) asymptotic analysis of Xu et al. [7]. We also plot experimental data of Xu et al. [7] using open and filled diamonds in Figure 8(b), representing $\Delta T < 0$ and $\Delta T > 0$, respectively. The present model predicts the sign of $\Delta T$ accurately against experimental data [7], as compared to the model of Xu et al. [7]. For instance, frame $h_S/R$ = 0.4 of Figure 8(b) shows that the experimental data for $\Delta T > 0$ is not accurately predicted by the model of Xu et al. [7] and our model predicts it accurately. Similarly, simulation data of Hu and Larson [5] is predicted well by the present model in Figure 8(b).

Interestingly, Figure 8(b) shows that area of the region under the curve of $\Delta T < 0$ obtained by the model of Xu et al. [7] increases with increase in $h_S/R$, as compared to that in the present model. In the present model, the area of the region under $\Delta T < 0$ decreases with increase in $h_S/R$. The trend of the present model is physically valid since, at larger $h_S/R$, a larger thermal energy is available and thereby, the threshold value of $k_S/k_d$ for the inversion of the sign of $\Delta T$ is lower.

## 4.2 Comparison between predictions of the model and measurements on a heated substrate

### 4.2.1 Time-averaged evaporation mass rate

We present comparisons of measured non-dimensional time-averaged evaporation mass rate ($\hat{\dot{m}}_{avg}$, eq. (12)) of evaporating water droplets on a heated substrate with predictions of two models. Two sets of measurements are utilized for the comparisons. The first set, reported by Sobac and Brutin [8] (mentioned as SB hereafter), considers evaporation of a 3.64 µL water droplet on a heated aluminum substrate. The wetted radius, initial contact angle and relative humidity are 1.44 mm, 68° and 0.475, respectively. For this case, the ratio of thermal conductivity of the substrate to that of the droplet ($k_S/k_d$) is 388. The second set, reported in the present work (section 3), considers evaporation of a 2.0 µL water droplet on a heated glass substrate. The wetted radius and initial contact angle are listed in Table 4 and $k_S/k_d = 1.5$ for glass substrate used in the present work.

In Figure 9, $\hat{\dot{m}}_{avg}$ increases non-linearly with substrate temperature ($T_S$) in the present as well as in the measurements of SB. The measured values in the present work are lower as compared to those of SB at a given $T_S$ because the latter study used a highly conductive substrate ($k_S/k_d = 388$) as compared to that in the present study ($k_S/k_d = 1.5$). In the latter study, the droplet receives larger thermal energy from the substrate and thereby the droplet evaporates faster, which results in larger $\hat{\dot{m}}_{avg}$.

The following models are tested for the comparison with the measurements mentioned above. The first model is reported by SB, which was built on the model of Hu and Larson [2]. The former model integrates the evaporation mass rate ($\dot{m}$) over total evaporation time ($t_F$) to obtain time-averaged evaporation mass rate ($\dot{m}_{avg}$, eq. (12)). In this model, the temperature of the liquid-gas interface is assumed to be equal to substrate temperature ($T_S$), and evaporative cooling due to the latent heat of evaporation along the liquid-gas interface is ignored. The second model is reported in the present work (section 2), which considers the evaporative cooling and variation of the diffusion coefficient of liquid vapor in air with temperature (eq. (7)). In the present model, we compute time-averaged evaporation mass rate ($\dot{m}_{avg}$), as suggested by SB and non-dimensionalize, as given by eq. (12).

In Figure 9, first three plots ($k_S/k_d = 388$, shown in red) shows comparison among variation of $\hat{\dot{m}}_{avg}$ with $T_S$ obtained in the measurements of SB and by predictions of two models mentioned above. Both models predict a non-linear increasing trend of $\hat{\dot{m}}_{avg}$ with $T_S$, similar to that reported in the measurement. We plot relative percentage errors in the models calculated with respect to the measurement in Figure 10(a). The error increases with $T_S$ for both models. The model of SB shows larger error as compared to the present model at a given $T_S$. The maximum errors for the former and latter are 30% and 19%, respectively, at $T_S = 75°C$. The increase in the error with respect to $T_S$ is attributed to larger convection in the droplet at larger $T_S$, which is ignored in both models. Since the present model shows significantly lesser error, it shows the importance of the following two effects, which were ignored in the model of SB. First, solving the energy equation in droplet and substrate allows to accurately estimate the liquid-gas interface temperature, I .e., evaporative cooling which occurs during the evaporation. Second, the dependence of diffusion coefficient of liquid vapor on temperature (eq. (7)) for a heated substrate.

In Figure 9, three plots ($k_S/k_d = 1.5$, shown in blue) compares the measured and calculated $\hat{\dot{m}}_{avg}$ as a function of $T_S$. The predictions of two previously-mentioned models are plotted, and the errors in two models are plotted in Figure 10(b). Comparison with predictions by the model of SB with present measurement shows that for $T_S = 50°C$ to $80°C$ and $T_S = 90°C$, the model underestimate and overestimate $\hat{\dot{m}}_{avg}$ respectively. While at $T_S < 50 °C$, the simulated values of $\hat{\dot{m}}_{avg}$ by both models are in good agreement with experiments, with an error of less than 10% (Figure 10(b)). At $50°C \leq T_S \leq 90°C$, the simulated values start deviating from the measured ones, and the absolute value of the maximum error is around 10%. Our numerical predictions underestimate the measurements since the model ignores the convection and considers only heat conduction. As mentioned earlier, a larger error is attributed to an intense convection inside the droplet at larger $T_S$.

The model of SB performs better than the present model for the present measurements ($k_S/k_d = 1.5$) for 80°C ≤ $T_S$ ≤ 90°C $T_S$ (Figure 10(b)), however, it underpredicts their own measurements ($k_S/k_d = 388$) for all cases of $T_S$ (Figure 9 and Figure 10(a)). This is explained as follows. The measurements of SB used a substrate of large thermal conductivity ($k_S/k_d = 388$) while a substrate of a lower thermal conductivity ($k_S/k_d = 1.5$) is used in the present work. Therefore, a larger thermal energy is available in the droplet from the substrate in the former and thereby resulting in a larger liquid-gas interface temperature (that is closer to substrate temperature) in the former as compared to the latter. A comparison of the simulated interface temperature by our model between the two cases at $T_S = 70°C$ is shown in Figure 10(c) and confirms this hypothesis. In SB model, the liquid-gas interface temperature is assumed to be equal to the substrate temperature (i.e. $T_{LG} = T_S$) and diffusion coefficient is taken at ambient temperature. The first and second assumption correspond to overprediction and underprediction of $\hat{\dot{m}}_{avg}$, respectively. In case of $k_S/k_d = 388$, $T_{LG}$ is closer to $T_S$ and the first assumption is valid, as explained above. However, SB model underpredicts measurements significantly for 40°C < $T_S$ < 90°C, due to the second assumption. In case of $k_S/k_d = 1.5$, the net effect of these two assumptions results in a better prediction by SB model at 80°C ≤ $T_S$ ≤ 90°C, as compared to the present model. The errors plotted in Figure 10(b) show that SB model overpredicts and underpredicts $\hat{\dot{m}}_{avg}$ for $T_S = 90°C$ and $T_S = 80°C$, respectively, showing that the first and second assumption becomes dominant, respectively.

Overall, $\hat{\dot{m}}_{avg}$ is overpredicted and underpredicted if liquid-gas interface temperature is taken equal to substrate temperature ($T_S$) and diffusion coefficient is not considered as a function of temperature, respectively. Therefore, these factors should be taken into account in the model in order to achieve reasonable fidelity. In particular, these two effects are more important for a low value of $k_S/k_d$ and a larger contact angle.

### 4.2.2 Time-varying droplet dimensions and temperature profile across liquid-gas interface

In this section, we compare numerical simulations and experimental data of droplet evaporation at different substrate temperatures. Time-varying volume and wetted diameter, and temperature at the edge and top of the droplet at $T_S = 27, 54$ and $91°C$ are compared. Figure 11(a) shows side view of droplet shapes at different times at $T_S = 91°C$. A horizontal line demarcates the reflection of the droplet in each frame. Figure 11(a) shows an almost pinned contact line during the evaporation and the droplet maintains a spherical cap shape, qualitatively confirmed by Figure 11(a). The spherical cap shape is attributed to the fact that the characteristic length (or wetted diameter) is lesser than the capillary length of water (~ 2.7 mm).

Figure 11(b), (c) and (d) show time-varying non-dimensional volume and wetted diameter at $T_S$ = 27, 54 and 91°C, respectively. The volume ($V$), wetted diameter ($d$) and time ($t$) are normalized with initial volume ($V_i$), initial wetted diameter ($d_i$) and total drying time ($t_F$), respectively. The expression of $t_F$ is given by Popov [4]:

$$t_F = \frac{\pi \rho R^2 \theta_i}{16 D_\infty (c_{sat} - H c_\infty)} \quad (14)$$

where $\theta_i$ and $c_{sat}$ are the initial contact angle and vapor concentration at substrate temperature, respectively. While calculating $t_F$ by eq. (14), we assume that the temperature of the liquid-gas interface equal to the substrate temperature ($T_S$). The model prediction, as well as measurement, shows a linear decrease of the volume with time in Figure 11(b-d). We also plot time-varying volume ($V(t)$) obtained by analytical model of Popov [4] and is given by,

$$V(t) = \frac{M(t)}{\rho} = \frac{\pi R^3 \theta_i}{4}\left(1 - \frac{t}{t_F}\right) \quad (15)$$

The maximum percentage error in the total evaporation time given by present model as well as the model of Popov [4] with respect to measurements is lesser than 15%. The latter model under-predicts $t_F$ because of $c_{sat}$ in eq. (14) is obtained at substrate temperature. The error in both models vary with substrate temperature and shows a similar trend as discussed in section 4.2.1 and as plotted in Figure 10(b).

Figure 12(a) shows the time-varying isotherms recorded from the top of a sessile droplet on a heated glass substrate by an infrared camera at $T_S$ = 91°C. The droplet reaches in thermal equilibrium with the substrate within 4 s, as noted from the isotherms. The droplet receives thermal energy from the substrate by heat conduction and loses heat at the liquid-gas interface due to the latent heat of evaporation. The heat conduction inside the droplet dominates over the convection due to smaller Péclet number, as explained earlier in section 2.1. Since the thermal conductivity of the substrate is larger than that of the droplet ($k_S/k_d$ = 1.5), the thermal energy is readily available near the contact line, and the temperature near the contact line region is larger than on the apex of the droplet [6]. The time-varying isotherms plotted at different instances attains a plateau value, as noted qualitatively in Figure 12(a).

Figure 12(b) and (c) plot the temperatures at the center ($T_{top}$) and the contact line (of the droplet for $T_S = 91^oC$ and $T_S = 54^oC$, respectively and compares them with the predictions of the present model. The measurements of both cases show that $T_{top}$ increases exponentially initially due to thermal equilibration in the droplet and attains a plateau value after the initial increase. The time-variation of $T_{top}$ is qualitatively confirmed by isotherms plotted in Figure 12(a) and can be explained by the thermal equilibration of the droplet with the heated substrate. On the other hand, $T_{edge}$ decays exponentially initially and attains a plateau value. $T_{edge}$ remains almost constant for most of the duration of the evaporation in Figure 12(b, c). This is explained as follows. At the onset of the evaporation, the temperature at the contact line ($T_{edge}$) is closer to the substrate temperature (as confirmed by Figure 12 (b, c) at $t = 0$ s). At $t > 0$, the droplet thermal equilibrates with the substrate and $T_{edge}$ decreases to a plateanu value. The plateau value of the temperature at the center of the droplet ($T_{top}$) is lesser than as compared to that at the contact line ($T_{edge}$) in both cases of the substrate temperature in Figure 12(b) and Figure 12(c) . The lower temperature at the center of the droplet ($T_{top}$) is attributed to a longer conduction path from the substrate to the top of the droplet as compared to that at the contact line. In the last stage of the evaporation, $T_{top}$ and $T_{edge}$ sharply increase in both cases in Figure 12(b, c) since the droplet becomes a thin liquid film at this time. In Figure 12 (b, c), the quasi-steady model does not capture the initial variation of $T_{top}$ and $T_{edge}$, however, the plateau values obtained in simulations in both cases of the substrate temperature are consistent with the measurements within a maximum error of 12%. A larger error at larger substrate temperature is attributed to a strong convection in the droplet, ignored in the model.

## 4.3  Numerical predictions of evaporation mass flux and evaporation mass rate on a heated substrate

Figure 13(a) and (b) plot computed non-dimensional evaporation mass flux ($\hat{j}$, eq. (9)) on the liquid-gas interface with respect to non-dimensional radial distances ($\hat{r} = r/R$) for different cases of substrate temperature ($T_S$) at contact angle ($\theta$) of 30° and 90°, respectively. In Figure 13(a), the spatial variation of $j$ at $T_S = 25^oC$ shows a larger value near the edge as compared to the top of the droplet, consistent with reported in previous studies (section 2.4). The profiles of $\hat{j}$ show the similar trend in cases of $T_S = 60°C$ and $90°C$, as plotted for $T_S = 25°C$, in Figure 13(a). We note that $\hat{j}$ increases with $T_S$ because of availability of larger thermal energy at a larger $T_S$. Comparing $\hat{j}$ at $\theta = 90°$ in Figure 13(b) with respective profiles at $\theta = 30°$ in Figure 13(a), we note that the values of $\hat{j}$ are lower for $\theta = 90°$ at a given $\hat{r}$ and $T_s$. This is because of the larger height of the droplet and thermal resistance for $\theta = 90°$, which lowers the temperature at the interface and thereby the value of $\hat{j}$. In order to quantify, the increase in the value of $\hat{j}$ at the edge ($\hat{j}_{edge}$) and at the top ($\hat{j}_{top}$) at a given $\theta$, we plot variation of $\hat{j}_{edge}$ and $\hat{j}_{top}$ against $T_S$ for different cases of $\theta$ in Figure 14(a) and Figure 14(b), respectively. In Figure 14(a), $\hat{j}_{edge}$ increases non-linearly with $T_S$, similar to increase in $c_{sat}$ with respect to $T_S$ (eq. (6)) in all cases of $\theta$. This is explained by the fact that the temperature at the edge is close to $T_S$. On the other hand, in Figure 14(b), $\hat{j}_{top}$ scales non-linearly at $\theta = 30°$ and this variation becomes almost linear at $\theta = 90°$. The value of $\hat{j}_{top}$ is significantly lesser at larger $T_S$ for larger $\theta$. This is explained by the larger thermal resistance offered by larger droplet height at larger $\theta$, which decreases interface temperature at the top of the droplet.

In order to quantify the effect of substrate temperature ($T_S$), the thickness of the substrate ($h_S$), substrate conductivity ($k_S$) and contact angle ($\theta$) on non-dimensional evaporation mass rate ($\hat{m}$, eq. (11)), we systematically vary these parameters and plot $\hat{m}$ with respect to $T_S$ in different frames of Figure 15. We select the following range of parameters in the simulations; $T_S = [25°C, 90°C]$, $h_S/R = [0.1, 10]$, $k_S/k_d = [1.5, 388]$ and $\theta = [10°, 90°]$. In all simulations, the droplet volume is kept same. The rows of Figure 15 show the simulations results for constant $k_S/k_d$ while columns show the simulations results for constant $h_S/R$. The top-left frame of Figure 15 shows a non-linear increase in $\hat{m}$ with respect to $T_S$ at $h_S/R = 0.1$ and $k_S/k_d = 1.5$ in three cases of contact angles ($\theta = 10°, 50°, 90°$). $\hat{m}$ for $\theta = 90°$ is slightly larger than $\theta = 10°$ because the wetted radius ($R$) for the former is smaller than the latter and $\hat{m}$ scales inversely with $R$ (eq. (11)). Therefore, the dimensional value of evaporation mass rate ($\dot{m}$) is smaller for $\theta = 90°$ as compared $\theta = 10°$. All remaining frames of Figure 15 show the similar trend of $\hat{m}$ with respect to $T_S$ at different $\theta$, as discussed above.

The first row of Figure 15 shows the effect of $h_S/R$ for a low conductive substrate ($k_S/k_d = 1.5$, representative of water and glass). As $h_S/R$ increases, $\hat{m}$ decreases for a given $T_S$ and $\theta$. In the third row of Figure 15, $\hat{m}$ remains unchanged with an increase in $h_S/R$ for $k_S/k_d = 388$ (representative of water and aluminum). In case of $k_S/k_d = 1.5$ (first row), as $h_S/R$ increases, the thermal resistance for the heat conduction from substrate bottom to the liquid-gas interface also increases, which hinders the heat transport. Due to this fact, for high $h_S/R$ as compared to low $h_S/R$, liquid-gas interface temperature is lower and therefore, the vapor concentration at the interface decreases, it results in lower $\hat{m}$. However, in case of $k_S/k_d = 388$, as $h_S/R$ increases, the transport of heat from substrate bottom to the liquid-gas interface does not get hindered due to lower thermal resistance as the thermal conductivity of the substrate is high. In this case, interface temperature is a weak function of $h_S/R$ and thereby $\hat{m}$ does not change appreciably.

The first column of Figure 15 shows the effect of $k_S/k_d$ for a thin substrate ($h_S/R = 0.1$) and $\hat{m}$ increases slightly with the increase of $k_S/k_d$ for a given $T_S$ and $\theta$. By contrast, this increase is larger for the third column ($h_S/R = 10$). This can be explained by the fact that, in case of $h_S/R = 0.1$ and a strong function in case of $h_S/R = 10$. In case of $h_S/R = 0.1$, the thermal resistance between substrate bottom to the liquid-gas interface is low for all cases of $k_S/k_d$, due to smaller heat conduction path from substrate bottom to the liquid-gas interface. However, in case of $h_S/R = 10$, a thicker substrate offers a larger thermal resistance, which decreases sharply as $k_S/k_d$ increases.

## 5 Conclusions

A combined numerical and experimental study is performed to study the effect of substrate heating on the evaporation of a sessile water droplet. We have developed a two-way coupled model for diffusion-limited, quasi-steady evaporation. The energy equation in the droplet and substrate and liquid-vapor diffusion equation in ambient surrounding the droplet were solved using the finite-element method in two-dimensional axisymmetric coordinates. Both equations were coupled by an iterative approach and under-relaxation is used to ensure numerical stability. The two-coupling is specifically important in case of larger contact angle, larger substrate temperature, lower thermally conductive substrate and a thicker substrate. The developed solver shows second-order accuracy and we have carried out validations of the evaporation mass flux and the temperature of the liquid-gas interface, reported in previous works. Our numerical model correctly predicts the inversion of the temperature gradient ($\Delta T$) across liquid-gas interface temperature, as a function of contact angle, substrate-droplet thermal conductivity ratio and substrate thickness-wetted radius ratio. The regimes of $\Delta T < 0$ and $\Delta T > 0$ were accurately predicted by the model as compared to previously reported one-dimensional model. We have employed high-speed visualization from the side and infrared thermography from the top of the evaporating droplet to measure time-varying droplet shapes and the temperature of the liquid-gas interface, respectively. The experiments were performed with water droplets on a heated glass substrate. Computed non-dimensional time-averaged evaporation mass rates ($\hat{\dot{m}}_{avg}$) at different substrate temperatures ($T_S$) were compared with the model and measurements of Sobac and Brutin [8] for a substrate with larger thermal conductivity (aluminum). We have also compared $\hat{\dot{m}}_{avg}$ computed by the present model and by the model of Sobac and Brutin [8], with present measurements for a substrate with a smaller thermal conductivity (glass). $\hat{\dot{m}}_{avg}$ increases nonlinearly with respect to $T_S$ in both cases of low and high thermally conductive substrate. Using these comparisons, we have highlighted the importance of two-way coupling used in the present model, which accommodates the evaporative cooling as well as the dependence of diffusion coefficient on temperature. We have compared the time-varying droplet volume and the liquid-gas interface temperature at the contact line and the top of the droplet, with numerical predictions obtained by the present model. The comparisons are in good agreement and validate the present model. The effect of the substrate heating on the evaporation mass flux is quantified and the flux near the contact line is found to increase by two to three orders of magnitude at $T_s > 50\ ^oC$. The flux is smaller for a larger contact angle due to significant evaporative cooling at the interface. We examine the combined effect of substrate heating, the ratio of substrate thickness to the wetted radius, substrate-droplet thermal conductivity ratio and the contact angle on the evaporation mass rate. The evaporation mass rate is not significantly influenced by the substrate thickness for a larger thermally conductive substrate. However, the evaporation mass rate inversely scales with the thickness in case of a lower thermally conductive substrate. In closure, the present paper provides insights on the relative importance of coupled transport phenomena as a function of associated parameters and these insights will be helpful to design engineering applications such as spray cooling and inkjet printing.

# 6  Acknowledgements

R.B. gratefully acknowledges the financial support of a CSR grant from Portescap Inc. India and of an internal grant from Industrial Research and Consultancy Centre (IRCC), IIT Bombay. M.K. thanks Mr. Nagesh D. Patil and Mr. Prathamesh G. Bange for assistance in performing the experiments.

# 8 Tables

Table 1: Definitions of notations used in the present paper

| Symbols | Definition |
| --- | --- |
| $A$ | liquid-gas interface area [m$^2$] |
| $c$ | liquid vapor concentration [kg/m$^3$] |
| $c_p$ | specific heat [J/kg K] |
| $d$ | wetted diameter of droplet [m] |
| $D$ | diffusion coefficient of liquid vapor in air [m$^2$/s] |
| $H$ | relative humidity [-] |
| $h$ | droplet height or substrate thickness [m] |
| $j$ | evaporative mass flux [kg/m$^2$ s] |
| $k$ | thermal conductivity [W/m-K] |
| $L$ | latent heat of evaporation [J/kg] |
| **n** | unit normal vector [-] |
| N | number of nodes [-] |
| $\dot{m}$ | evaporation rate [kg/s] |
| $M$ | mass [kg] |
| $r$ | radial coordinate [m] |
| $R$ | wetted radius of droplet [m] |
| $t$ | time [s] |
| $T$ | temperature [°C] |
| $V$ | volume [m$^3$] |
| $z$ | axial coordinate [m] |
| *Greek letters* | |
| $\alpha$ | thermal diffusivity [m$^2$/s] |
| $\varepsilon$ | $L_2$-norm error |
| $\theta$ | contact angle [radians or degrees] |
| $\rho$ | density [kg/m$^3$] |

*Superscript*

^          Non-dimensional value

*Subscripts*

*avg*       average

∞          ambient

*d*         droplet

*F*         total evaporation time of droplet

*i*         initial condition

LG         liquid-gas interface

*ref*       reference value

*s*         substrate

*sat*       saturated

*z*         axial direction

Table 2: Thermophysical properties (at 25°C) used in the simulations

| Substance | Density (kg/m$^3$) | Thermal conductivity (W/m K) | Specific heat (J/kg K) | Latent heat (KJ/kg) |
|---|---|---|---|---|
| Water | 1000 | 0.61 | 4187 | 2264 |
| Glass | 2600 | 0.96 | 840 | - |
| Aluminum | 2700 | 237 | 900 | - |

Table 3: Number of nodes in different grids utilized in grid-size convergence study.

| Cases | Nodes in droplet and substrate | Nodes in ambient |
|---|---|---|
| Grid 1 | 2789 | 2312 |
| Grid 2 | 6846 | 6913 |
| Grid 3 | 13461 | 10682 |
| Grid 4 | 17769 | 15028 |
| Grid 5 | 22851 | 23390 |

Table 4: Measured initial wetted radius and the initial contact angle of water droplets on a heated glass at different temperatures.

| Substrate temperature ($^oC$) | Wetted radius (mm) | Initial contact angle |
|---|---|---|
| 27 | 1.7 | 28.6º |
| 48 | 1.4 | 44.0º |
| 54 | 1.7 | 31.6º |
| 71 | 1.7 | 27.9º |
| 81 | 1.8 | 27.4º |
| 91 | 1.7 | 31.5º |

# 9 Figures

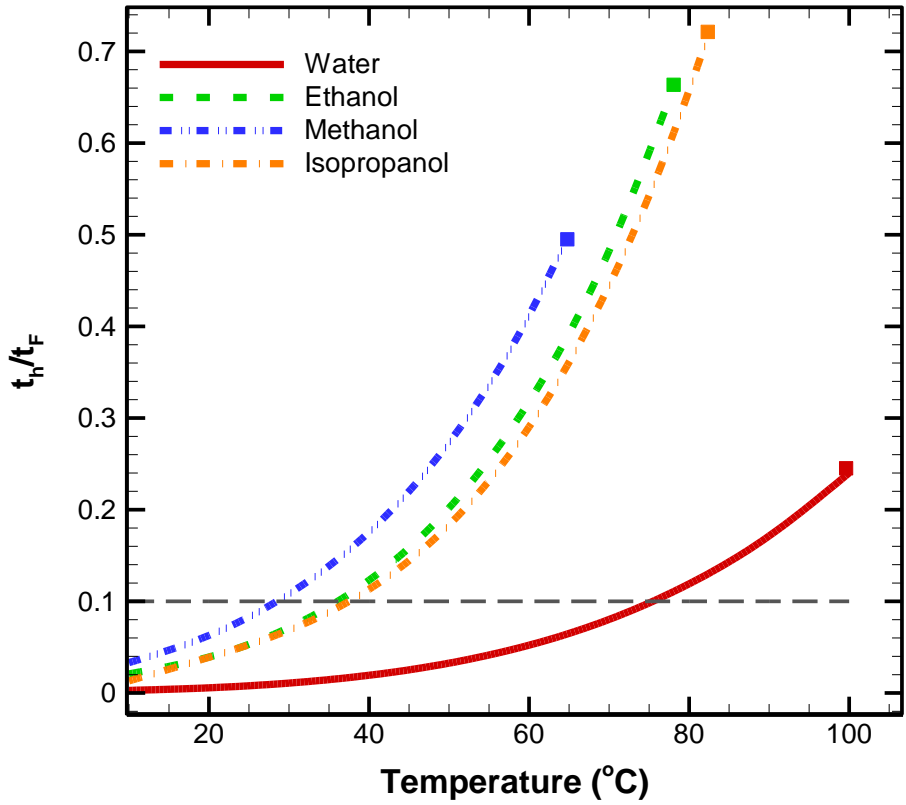

Figure 1: Ratio of heat equilibrium time in a droplet ($t_h$) and its total evaporation time ($t_F$) as a function of the average temperature of the droplet. A filled square on each curve represents the boiling point of the respective liquid. A horizontal dashed line represents $t_h/t_F = 0.1$, and assumption of quasi-steady evaporation is valid for $t_h/t_F < 0.1$, as suggested by Larson [24].

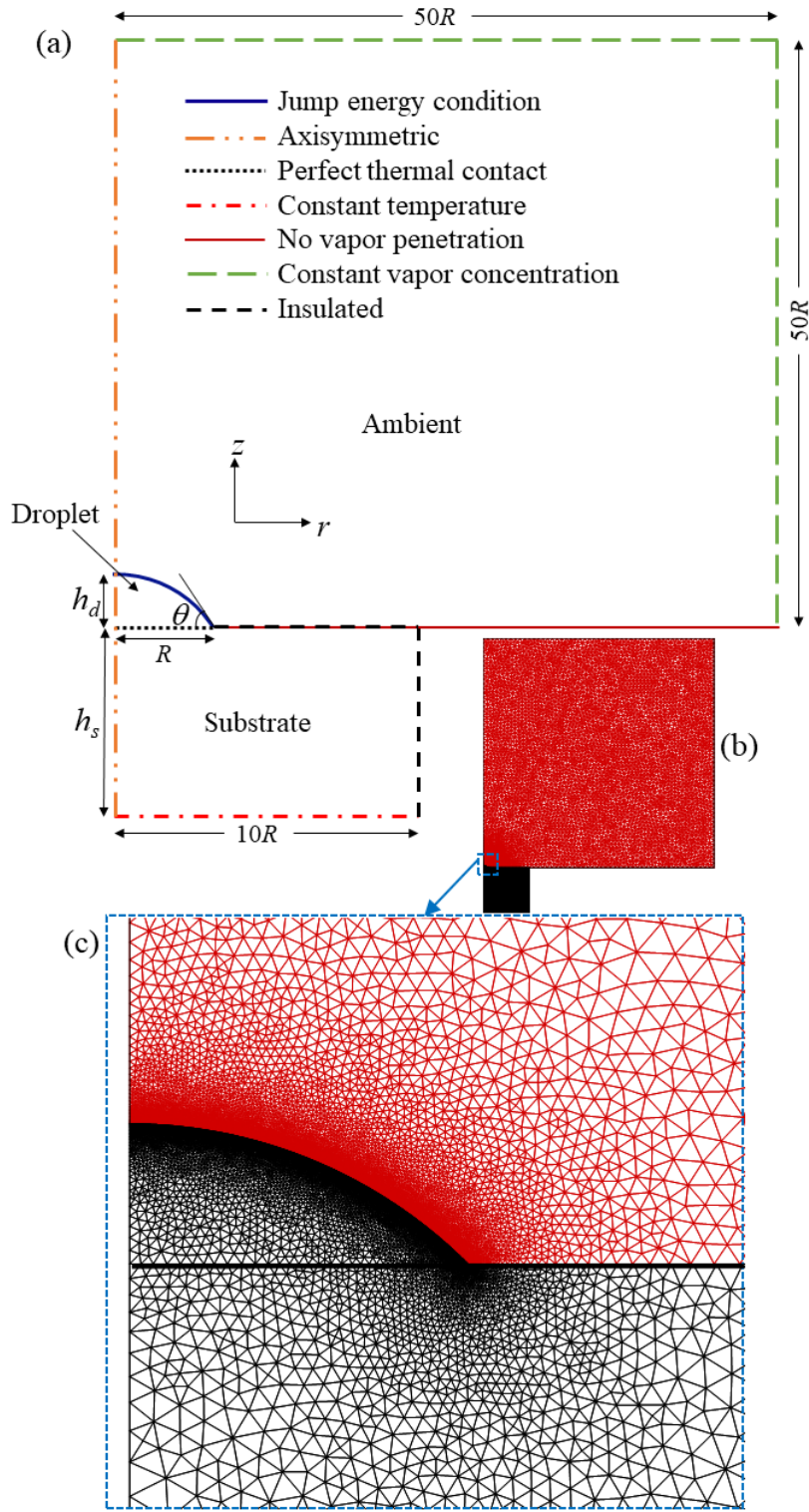

Figure 2: (a) Computational domain (not to scale) and boundary conditions used in the model, (b) A typical finite element mesh used in the simulations, (c) Zoomed-in view of the mesh in the droplet and in surrounding region.

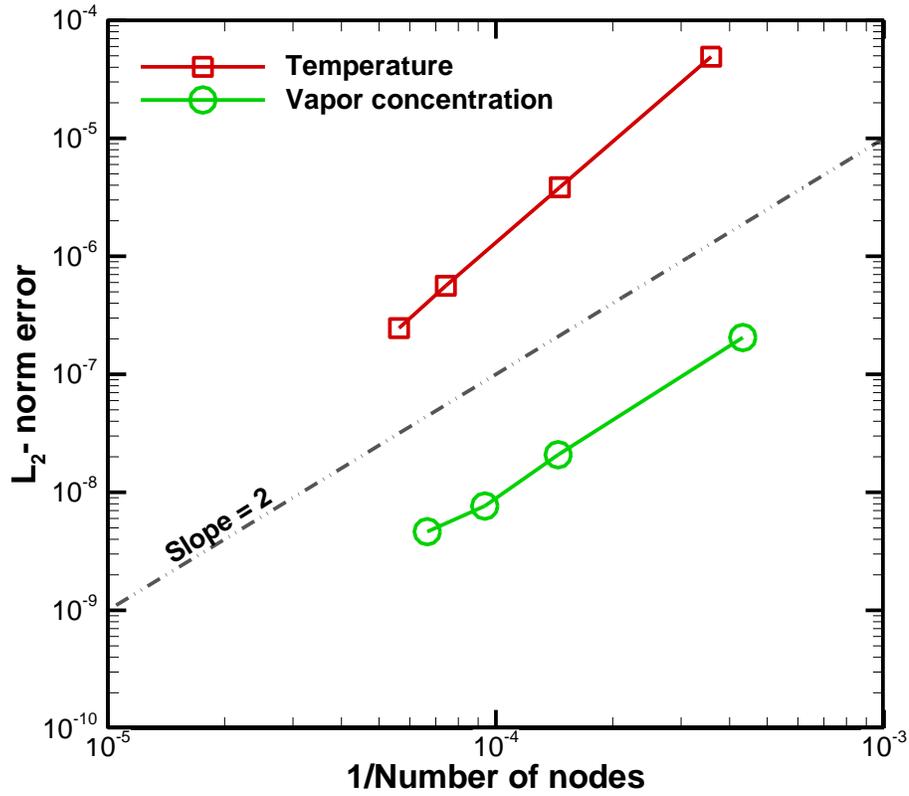

Figure 3: Grid size convergence study. $L_2$-norm error in temperature and vapor concentration solution is plotted with respect to different grids tested. A dashed line of slope 2 is plotted for reference. The present model shows second-order accuracy in the solution.

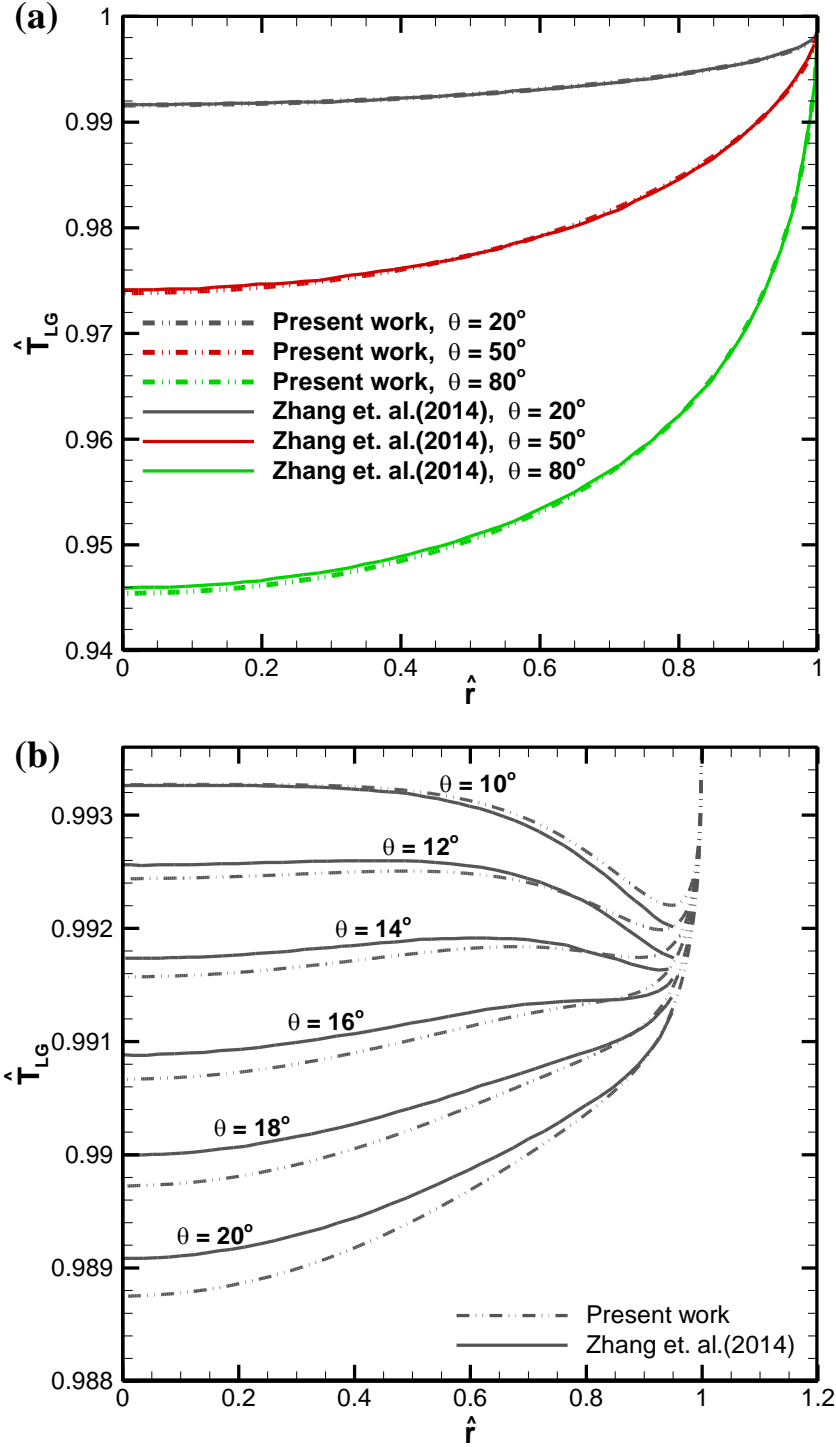

Figure 4: Non-dimensional temperature of the liquid-gas interface ($\hat{T}_{LG}$) with respect to non-dimensional radial distance ($\hat{r}$). The corresponding profiles obtained by Zhang et al. [9] are also plotted for comparison. (a) Substrate-droplet thermal conductivity ratio ($k_S/k_d$) and substrate thickness-wetted radius ratio ($h_S/R$) are 10 and 0.3, respectively. (b) $k_S/k_d = 1.58$, $h_S/R = 0.15$.

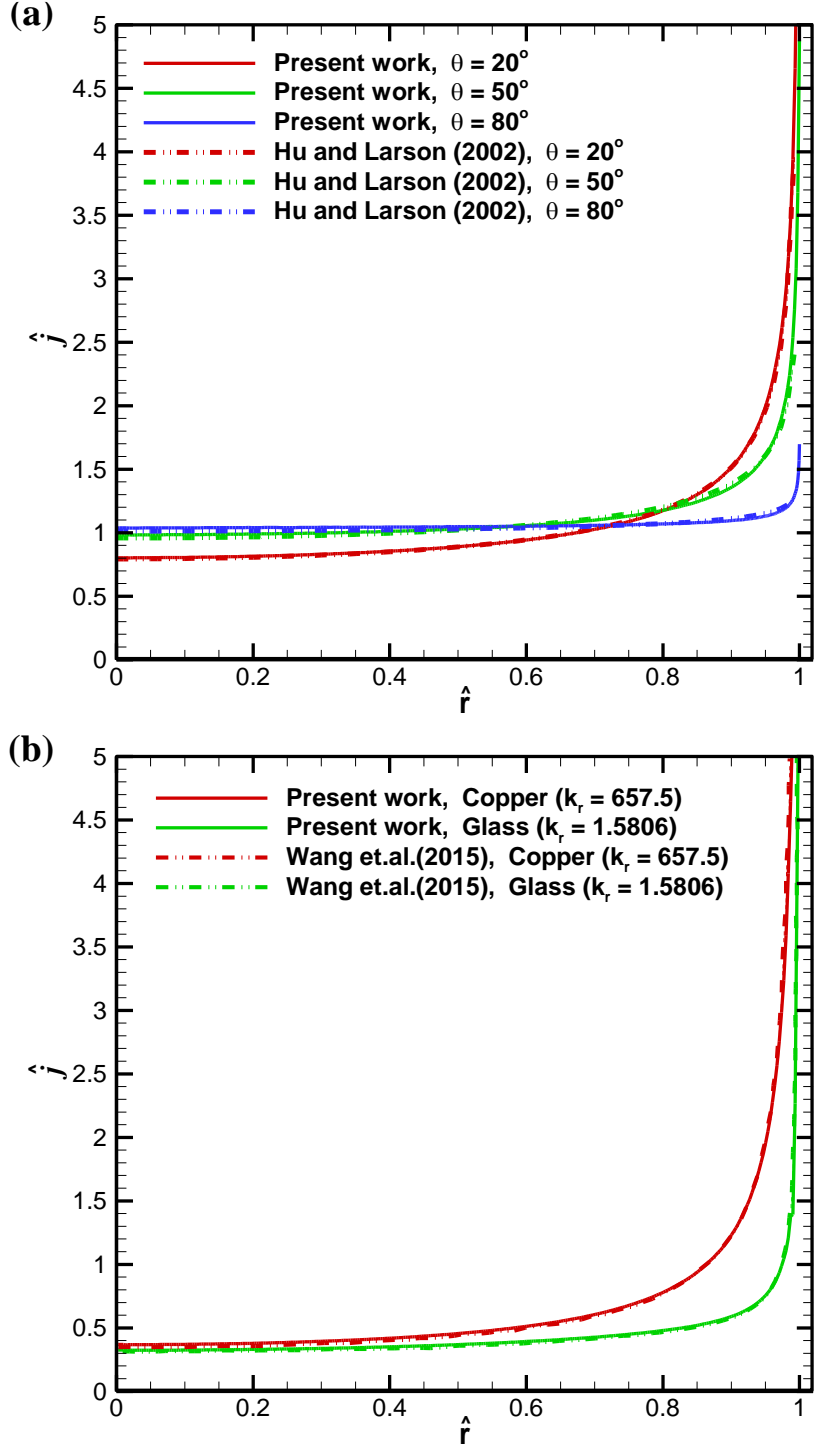

Figure 5: Non-dimensional evaporation mass flux at the liquid-gas interface ($\hat{j}$) with respect to non-dimensional radial distance ($\hat{r}$). The corresponding profiles obtained by Hu and Larson [2] and Wang et al. [13] are also plotted for comparison in (a) and (b), respectively. (a) Substrate-droplet thermal conductivity ratio ($k_S/k_d$) and substrate thickness-wetted radius ratio ($h_S/R$) are 10 and 0.3, respectively, (b) $h_S/R = 0.15$ and contact angle ($\theta$) = 10°.

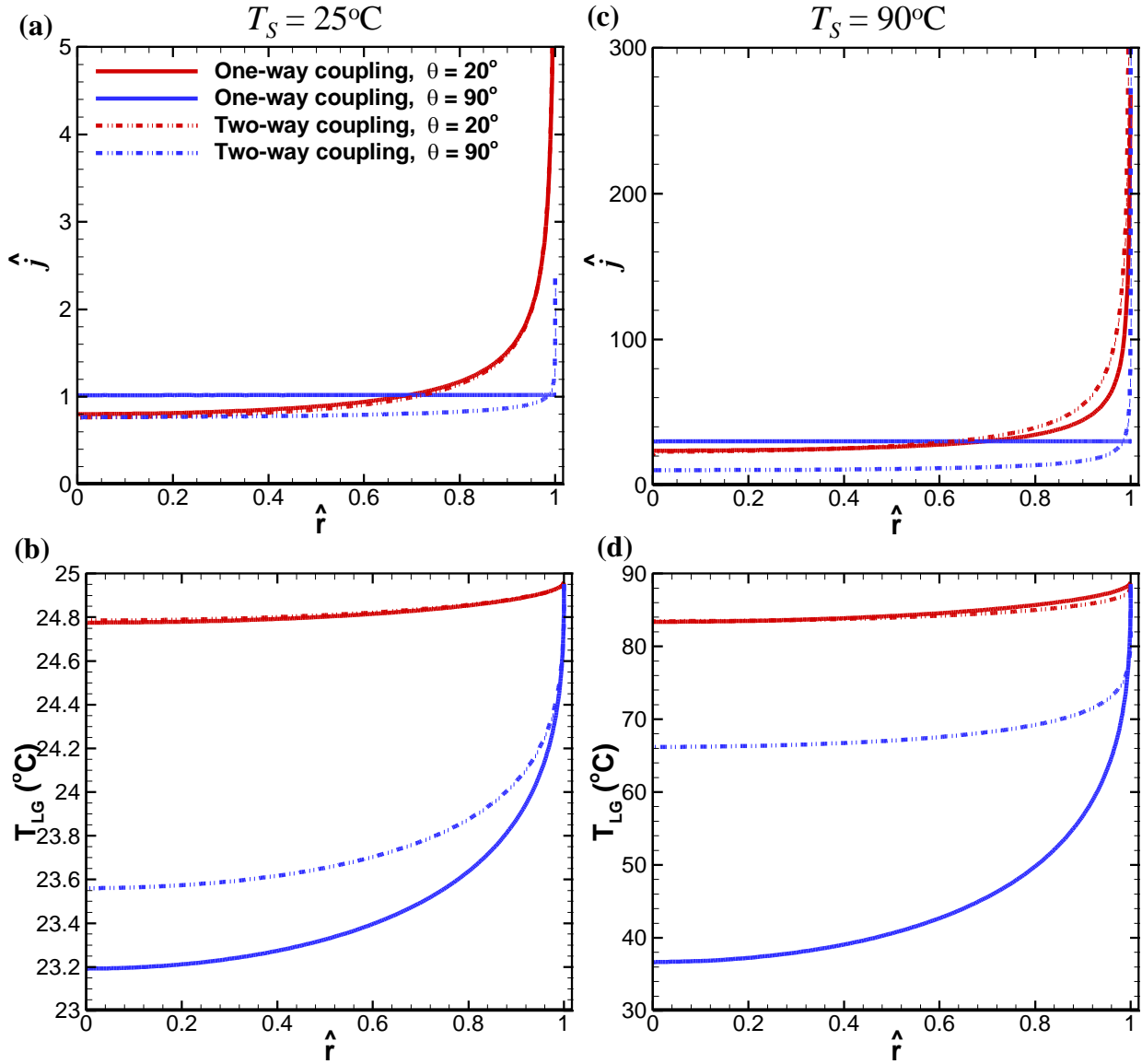

Figure 6: Comparison of variation of non-dimensional evaporation mass flux ($\hat{j}$) and liquid-gas interface temperature ($T_{LG}$) with respect to radial distance ($\hat{r}$), predicted by one-way and two-way coupling models for $\theta = 20°$ and $\theta = 90°$ at substrate temperature, $T_s = 25°C$ (a, b), and $T_s = 90°C$ (c, d). Substrate-droplet thermal conductivity ratio ($k_S/k_d$) and substrate thickness-wetted radius ratio ($h_S/R$) are 10 and 0.3, respectively.

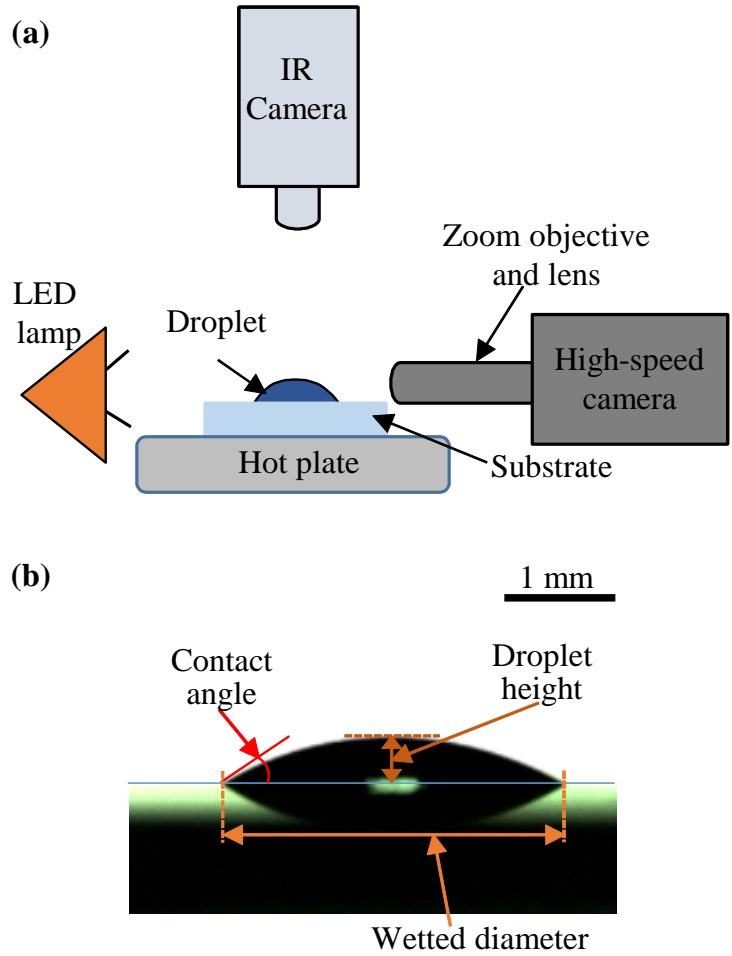

Figure 7: (a) Schematic illustration of the experimental setup, (b) Droplet dimensions are shown in an image recorded from side view by the high-speed camera.

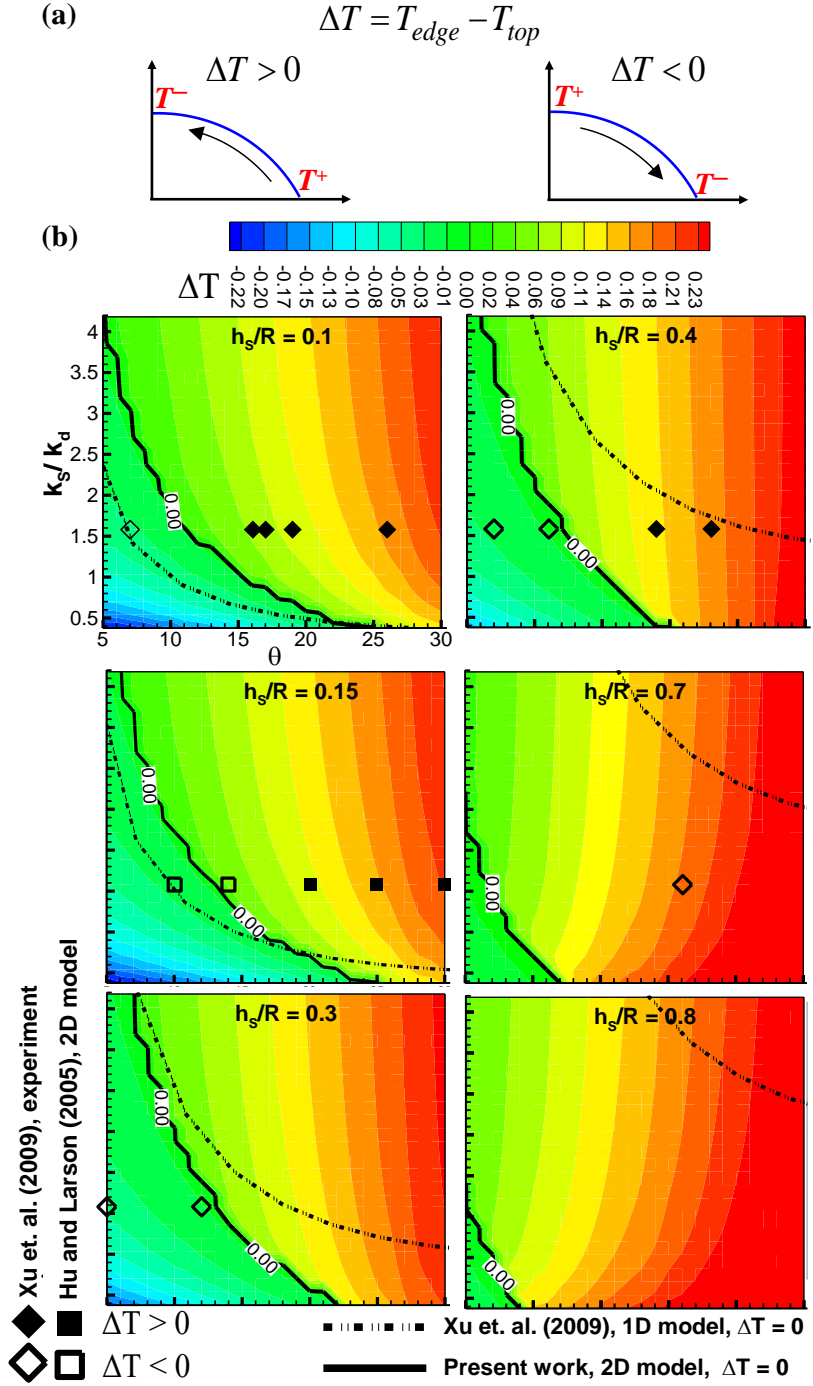

Figure 8: (a) Schematic illustration of positive ($\Delta T > 0$) and negative ($\Delta T < 0$) temperature difference along the liquid-gas interfaces shown in the left and right frame, respectively. The direction of flow along interface caused by thermal Marangoni stresses is also shown. (b) Computed contours of $\Delta T$ plotted on contact angle ($\theta$) - substrate-droplet thermal conductivity ratio ($k_S/k_d$) plane. Results obtained for different cases of the ratio of substrate thickness and wetted radius ($h_S/R$) are plotted in different frames. Broken and solid curve corresponding to $\Delta T = 0$ obtained from a model of Xu. et al. [7] and of present work, respectively. Filled and hollow symbols correspond to $\Delta T > 0$ and $\Delta T < 0$, respectively.

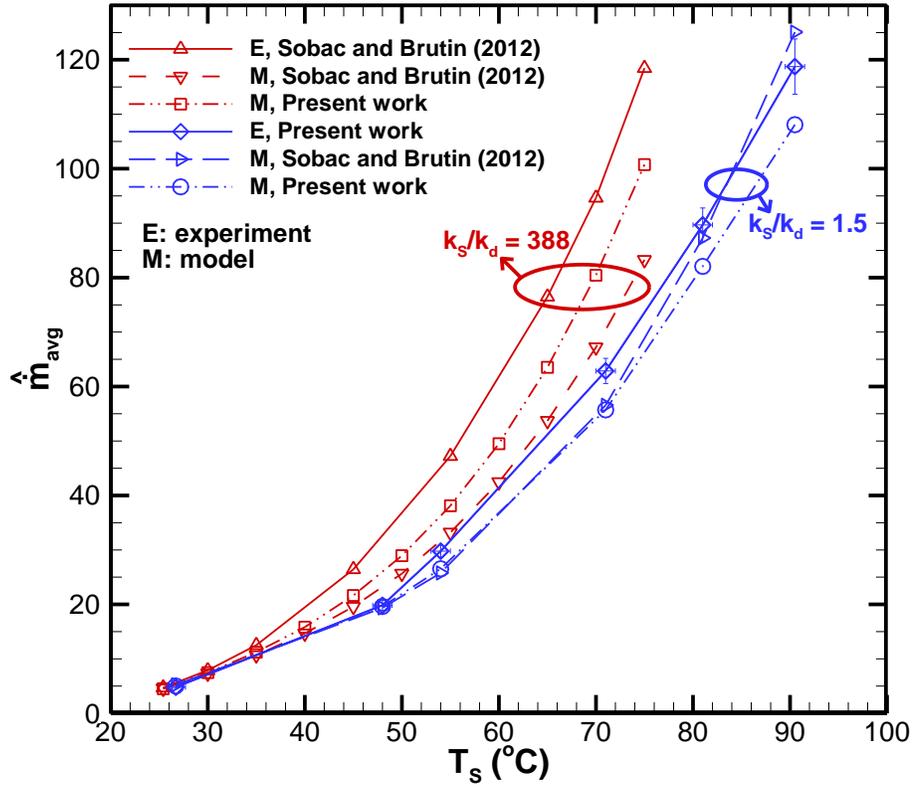

Figure 9: Comparison between experimental data and model predictions for dimensionless time-averaged evaporation mass rate ($\hat{\dot{m}}_{avg}$) against substrate temperature ($T_S$). Two sets of measurements are plotted for different cases of the ratio of thermal conductivity of the substrate to that of the droplet: $k_S/k_d = 388$ (red triangles) and 1.5 (blue diamonds), corresponding to experiments of Sobac and Brutin [8] and of the present work, respectively. These sets are compared with the numerical models of Sobac and Brutin [8] and of the present study.

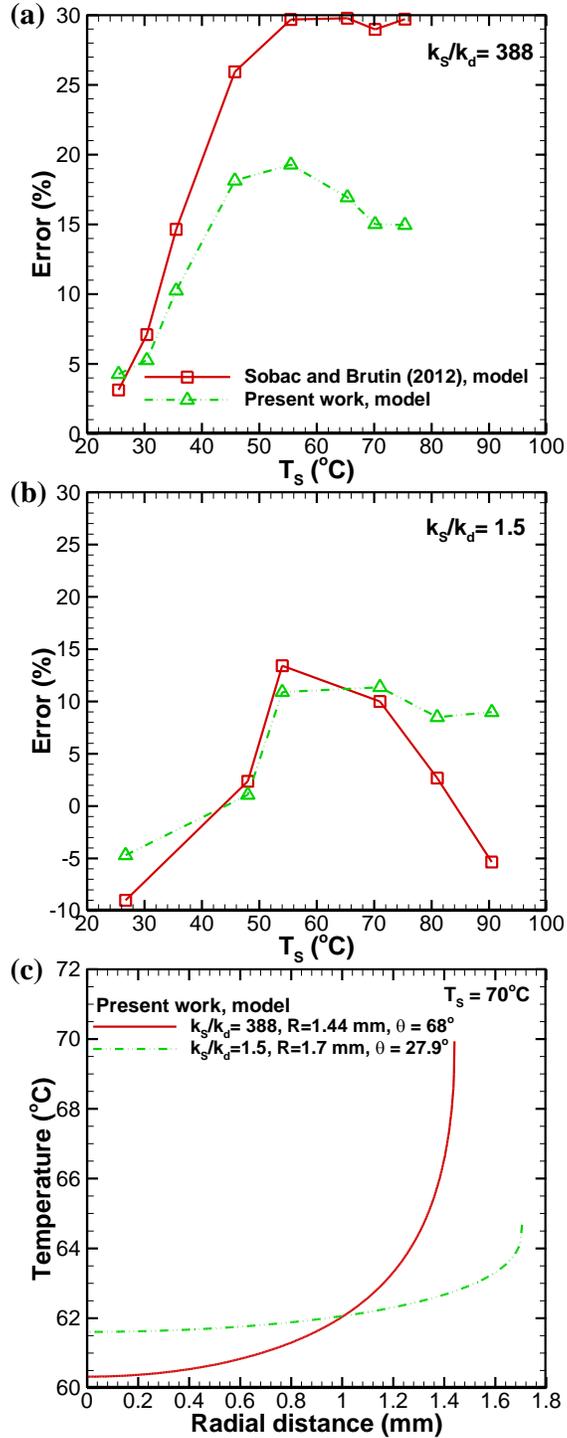

Figure 10: Percentage error in the model of Sobac and Brutin [8] and of the present work with respect to experiments of (a) Sobac and Brutin [8], $k_S/k_d = 388$ and (b) of the present work, $k_S/k_d = 1.5$. (c) Comparison between computed spatial variation of the temperature of the liquid-gas interface obtained for $k_S/k_d = 1.5$ and $k_S/k_d = 388$ for $T_S = 70^\circ C$.

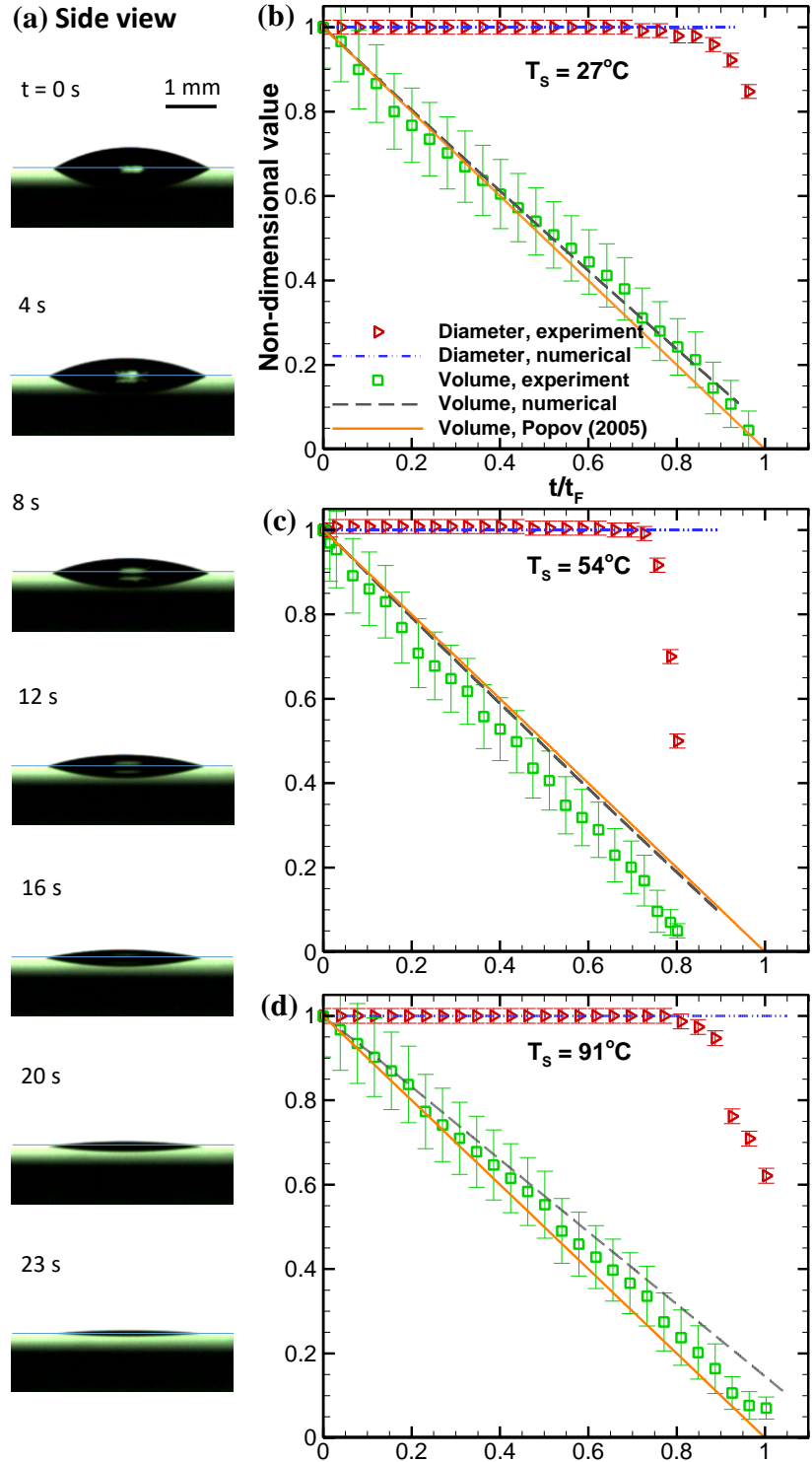

Figure 11: (a) Time-varying droplet shapes recorded in side-view by the high-speed camera for substrate temperature, $T_S = 91°C$. (b), (c) and (d) Comparison between computed and measured time-varying non-dimensional volume ($V/V_i$) and wetted diameter ($d/d_i$) for $T_S = 27°C$, 54°C and 91°C, respectively. The time is non-dimensionlized with respect to total evaporation time ($t_F$).

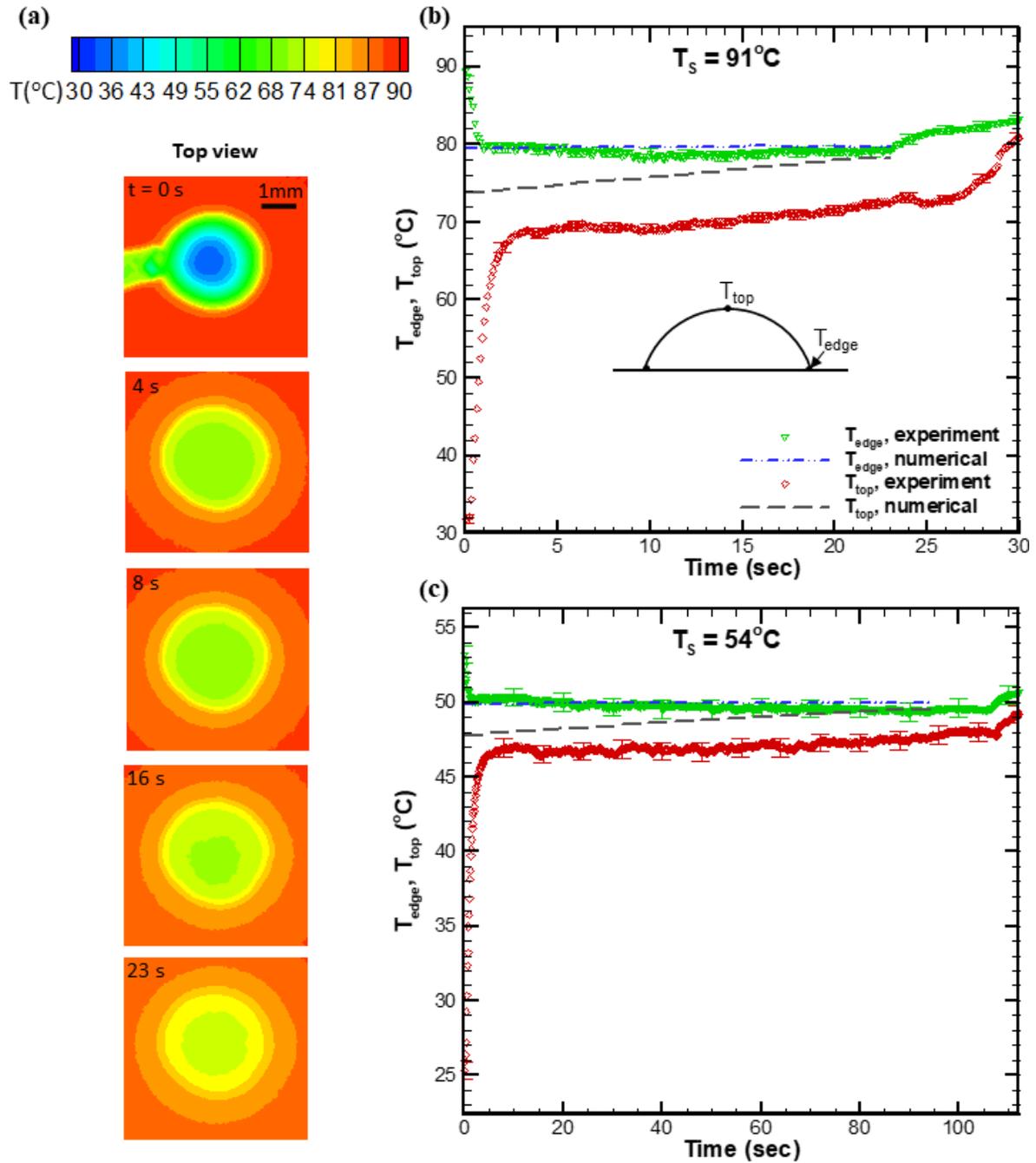

Figure 12: (a) Time-varying isotherms recorded in top-view by an infrared camera for substrate temperature, $T_S$ = 91°C. (b) and (c) Comparison of computed temperature at droplet edge ($T_{edge}$) and at the top ($T_{top}$) with respective measurements for $T_S$ = 54 and 91°C, respectively.

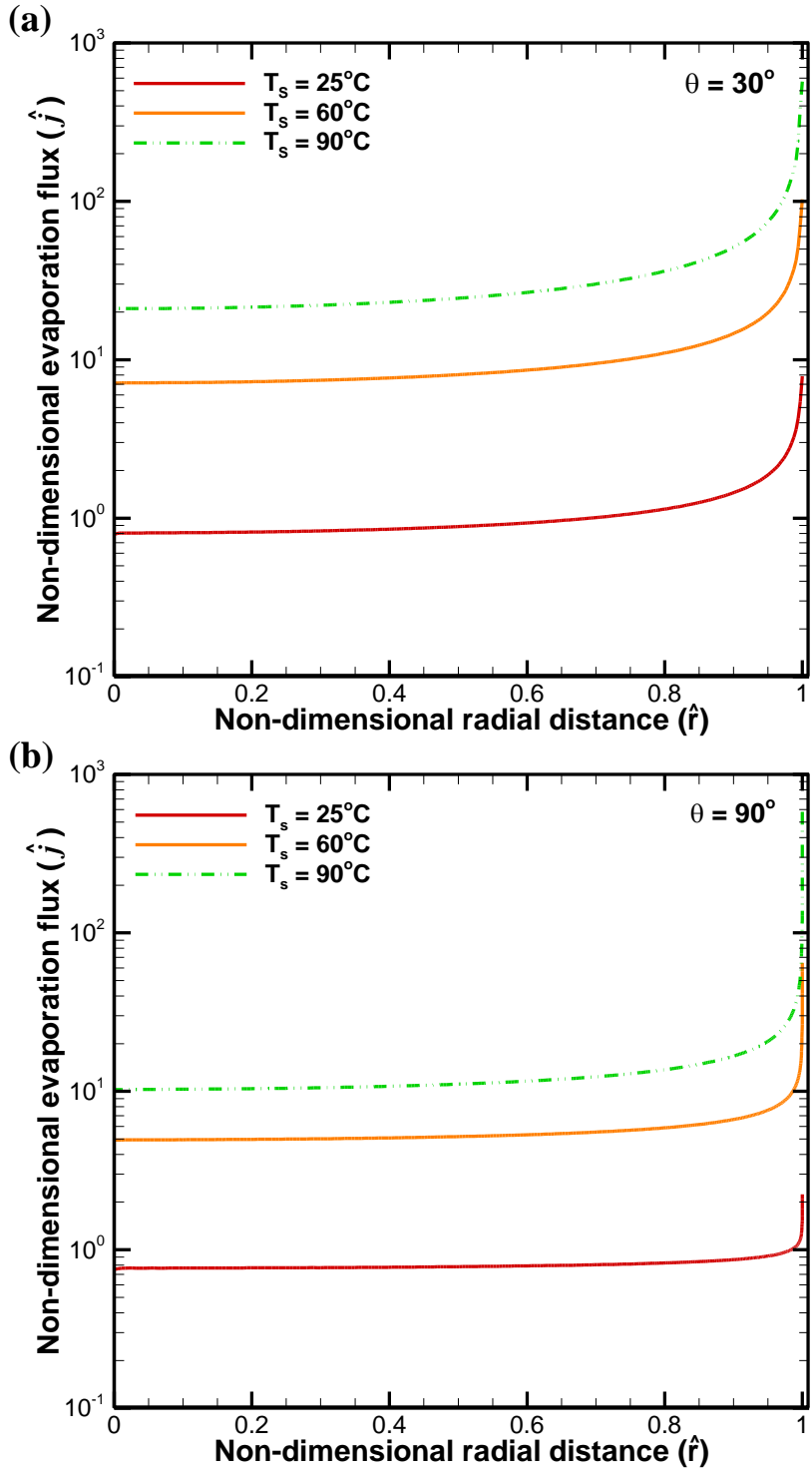

Figure 13: Variation of non-dimensional evaporation mass flux ($\hat{j}$) with respect to non-dimensional radial distance ($\hat{r}$) at different substrate temperatures, $T_S$ = 25, 60 and 90°C. $\hat{j}$ is plotted on log scale (a) Contact angle ($\theta$) = 30° and, (b) $\theta$ = 90°. Substrate-droplet thermal

conductivity ratio ($k_S/k_d$) and substrate thickness-wetted radius ratio ($h_S/R$) are 10 and 0.3, respectively.

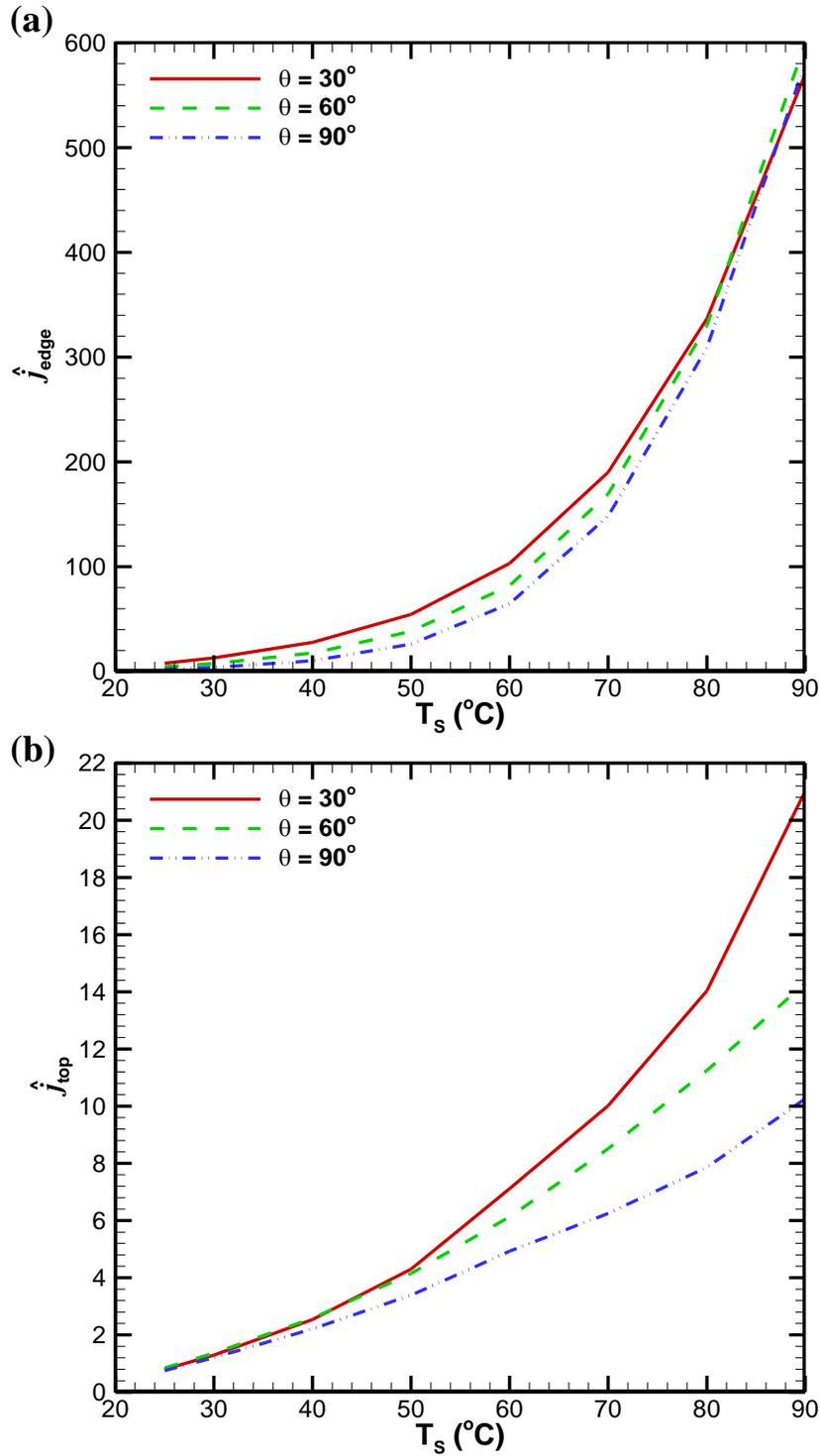

Figure 14: Non-dimensional evaporation mass flux at (a) edge ($\hat{j}_{edge}$) and (b) top ($\hat{j}_{top}$) with respect to substrate temperature ($T_S$). Three cases of contact angles, $\theta = 30°$, $60°$ and $90°$, are considered.

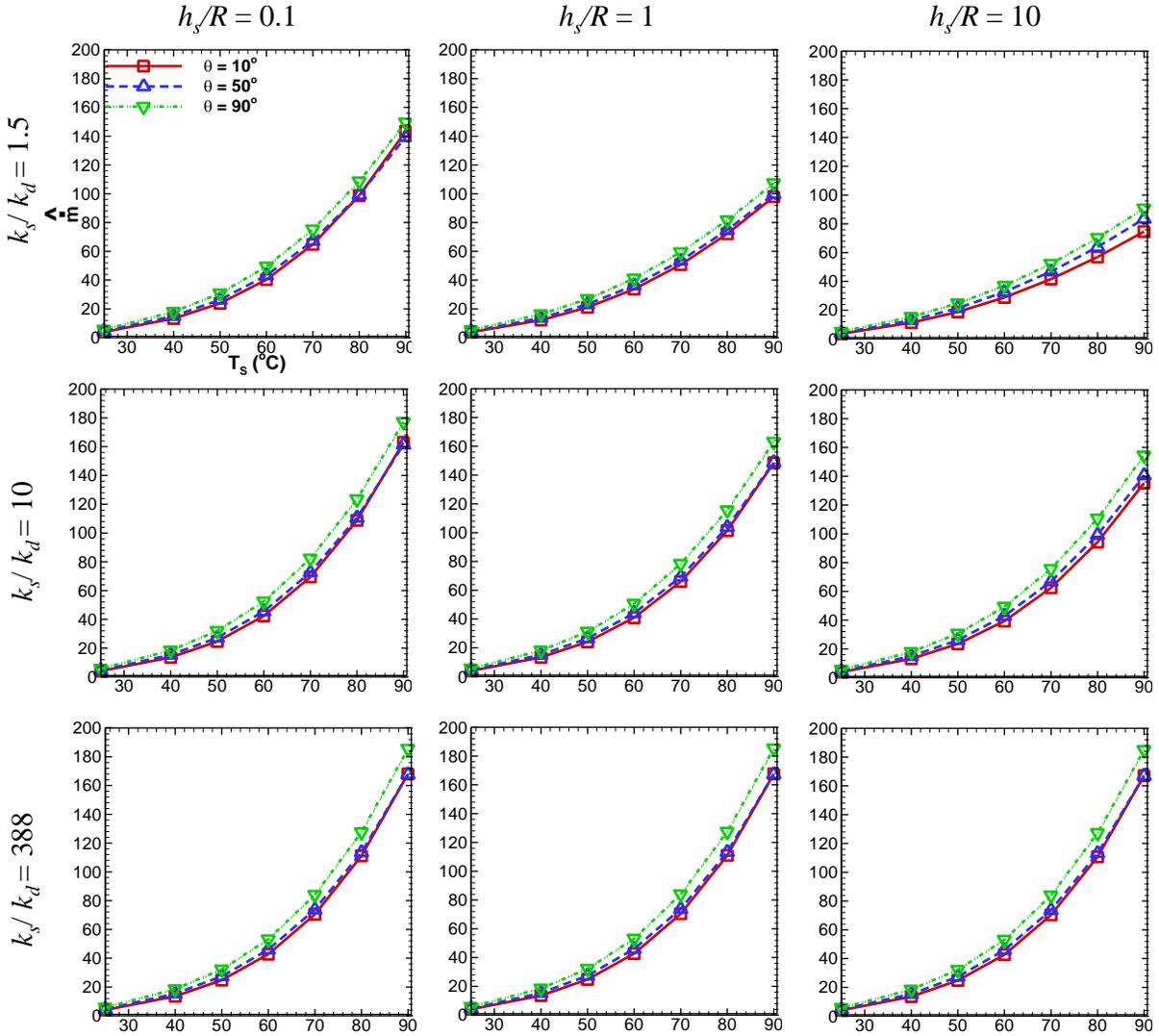

Figure 15: Effect of substrate thickness-wetted radius ratio ($h_S/R$), substrate-droplet conductivity ratio ($k_S/k_d$) and contact angle ($\theta$) on the variation of non-dimensional evaporation mass rate ($\hat{\dot{m}}$) with respect to substrate temperature ($T_S$).

43